\documentclass[pra,twocolumn,superscriptaddress,notitlepage,floatfix]{revtex4-2}

\usepackage[T1]{fontenc}

\usepackage{graphicx}
\usepackage{dcolumn}
\usepackage{bm}
\usepackage[colorlinks=true,urlcolor=blue,citecolor=blue,anchorcolor=red]{hyperref}
\usepackage{booktabs}

\usepackage{physics}
\usepackage{xcolor}
\usepackage{dsfont}
\hypersetup{
    colorlinks=true,
    linkcolor=blue,
    urlcolor=blue,
    citecolor=blue,
    }
\usepackage{soul}
\usepackage[version=4]{mhchem} 
\usepackage[T1,T2A]{fontenc}
\usepackage[normalem]{ulem}

\begin{document}

\title{Noisy intermediate-scale quantum simulation of the one-dimensional wave equation}

\author{Lewis Wright}
\email{lewis.wright@quantinuum.com}
\affiliation{Quantinuum, Partnership House, Carlisle Place, London SW1P 1BX, United Kingdom}
\author{Conor Mc Keever}
\affiliation{Quantinuum, Partnership House, Carlisle Place, London SW1P 1BX, United Kingdom}
\author{Jeremy T. First}
\affiliation{bp, Center for High Performance Computing, 225 Westlake Park Blvd, Houston, Texas 77079, USA}
\author{Rory Johnston}
\affiliation{bp, Center for High Performance Computing, 225 Westlake Park Blvd, Houston, Texas 77079, USA}
\author{Jeremy Tillay}
\affiliation{bp, Center for High Performance Computing, 225 Westlake Park Blvd, Houston, Texas 77079, USA}
\author{Skylar Chaney}
\affiliation{bp, Digital Science and Engineering, 501 Westlake Park Blvd, Houston, Texas 77079, USA}
\author{Matthias Rosenkranz}
\affiliation{Quantinuum, Partnership House, Carlisle Place, London SW1P 1BX, United Kingdom}
\author{Michael Lubasch}
\affiliation{Quantinuum, Partnership House, Carlisle Place, London SW1P 1BX, United Kingdom}

\begin{abstract}
We design and implement quantum circuits for the simulation of the one-dimensional wave equation on the Quantinuum H1-1 quantum computer.
The circuit depth of our approach scales as $O(n^{2})$ for $n$ qubits representing the solution on $2^{n}$ grid points, and leads to infidelities of $O(2^{-4n} t^{2})$ for simulation time $t$ assuming smooth initial conditions.
By varying the qubit count we study the interplay between the algorithmic and physical gate errors to identify the optimal working point of minimum total error.
Our approach to simulating the wave equation can be used with appropriate state preparation algorithms across different quantum processors and serve as an application-oriented benchmark.
\end{abstract}

\maketitle

\section{Introduction}

Partial differential equations (PDEs) are fundamentally important in scientific computing.
They play a crucial role in describing and understanding physical phenomena across a range of scales and scientific disciplines, including seismology~\cite{li2022research}, fluid dynamics~\cite{zawawi2018review}, financial mathematics~\cite{campolieti2018financial}, chemistry~\cite{pople1999nobel} and quantum mechanics~\cite{mcardle2020quantum}.

In this article we focus on the wave equation, whose solutions span fluid dynamics~\cite{durran2013numerical, alquran2012solitary}, electromagnetism~\cite{becherrawy2013electromagnetism, chew2022integral} and quantum chemistry~\cite{broglie1924xxxv,schrodinger2003collected, pople1999nobel}.
In particular, the wave equation is paramount to the energy industry for the exploration for hydrocarbons, discovery of shallow depth hazards and monitoring sequestered \ce{CO2}.
This hyperbolic PDE governs the time evolution of a pressure front propagating through a medium whose properties are defined by its acoustic velocity and is the primary compute kernel for a range of applications in seismic imaging~\cite{virieux2009fwi, virieux2017introduction}.
Indeed, the modeling of wave propagation is a large consumer of high-performance computing resources worldwide~\cite{brandsberg2017high}.
For this reason, algorithms to simulate the wave equation make excellent benchmarks for evaluating new classical computing architectures based on CPUs and GPUs, as well as emergent technologies such as quantum computing.

For the purpose of velocity model building in seismic imaging, the wave equation is cast as a PDE within an optimization problem that compares simulated seismic traces to experimental measurements after injection of a source wave.
In this method, known as full-waveform inversion~\cite{virieux2009fwi, virieux2017introduction}, a trial subsurface model is constructed and the optimal subsurface model parameters are found via gradient-based methods that convolve the forward and backward wavefield traveling through the trial subsurface.
Due to its outsized reliance on the acoustic wave equation, full-waveform inversion is fundamentally limited by the computational expense of modeling wave propagation through the subsurface.

While many formulations of the wave equation exist, the simplest form is the isotropic acoustic wave equation
\begin{equation}\label{eq: acoustic wave equation}
 \frac{\partial^2}{\partial t^2} \boldsymbol{u} = c^2 \Delta \boldsymbol{u},
\end{equation}
where $\boldsymbol{u}$ is the acoustic wavefield, $c$ is the velocity of the wavefield through a medium that is independent of direction and $\Delta$ represents the Laplacian, a second-order differential operator.

Numerical approaches to model wave propagation on classical hardware can broadly be classified into either spatial finite difference methods or spectral methods.
While spatial finite difference methods discretize space into a finite grid, spectral methods discretize the frequency spectrum; both methods discretize time into explicit time steps~\cite{etgen1989accurate, etgen2007computational, fornberg1986seg}.
However, whereas spatial finite difference methods calculate gradients across a finite convolutional derivative stencil in the spatial domain, spectral methods convert the spatial domain to the Fourier domain, apply the derivative operator and then convert back to the spatial domain.
Generally, spatial finite difference methods require higher-resolution grids to achieve the same accuracy as spectral methods because of the dispersion that occurs due to the spatial discretization.
On the contrary, spectral methods incur a computational overhead due to the forward and backward Fourier transforms.
In addition, the Fourier transform requires all-to-all communication, limiting the ability of the spectral method to be efficiently distributed across a high-performance computing cluster.
Nevertheless, the spectral method remains advantageous for many applications due to the lower numerical dispersion, allowing the use of lower spatial resolution grids for similar accuracy.

Like their classical counterparts, quantum algorithms for simulating PDEs on quantum computers typically aim to prepare solutions by discretizing time and/or space.
A key property of quantum algorithms is the ability to encode the discretized solution in the amplitudes of a quantum state and thereby achieve exponential memory compression~\cite{ZalkaSimulating1998, ChuangQuantum2010, childs2022quantum}.
By contrast, classical methods scale linearly with memory, meaning a 3D grid of twice the resolution requires $\times$ 8 the memory ($\times$ 2 for each dimension).
Given the slow growth of accessible memory, this imposes a classical ceiling for high resolution wave equation modeling.
A variety of quantum algorithms have been proposed to simulate both linear and non-linear PDEs.
These include algorithms based on solving linear equations~\cite{Cao2013Quantum,berry2014high,Berry2017Quantum,childs2017quantum,lloyd2020quantum,childs2021high, Krovi2023improvedquantum,liu2021efficient}, Hamiltonian simulation~\cite{Wiesner1996Simulations,ZalkaSimulating1998,costa2019quantum,arrazola2019quantum,jin2022quantum, babbush2023} and quantum phase estimation~\cite{XuEtAl18, XuEtAl19}, see also~\cite{GiviEtAl20}.
While many of these algorithms offer impressive performance guarantees, they can lead to deep quantum circuits requiring fault tolerant quantum computers.
Variational approaches for simulating PDEs \cite{lubasch2020,liu2021,sato2021,kyriienko2021solving,joo2021quantum,budinski2021quantum, albino2022solving,Guseynov2023Depth,liu2023variational,JakschEtAl23} can operate on current Noisy Intermediate-Scale Quantum (NISQ) devices \cite{preskill2018quantum}, but are typically heuristic in nature and do not offer asymptotic performance guarantees.

One quantum algorithmic approach for simulating the wave equation by Costa et al.~\cite{costa2019quantum} recasts the problem in terms of the evolution of the time-dependent Schr\"{o}dinger equation.
This work focuses on the implementation of complicated geometries with Dirichlet and Neumann boundary conditions and constructs a quantum algorithm in terms of the graph Laplacian.
For evolution time $t$, spatial dimension $d$, domain size $l$ per dimension and lattice spacing $a$, they implement the time evolution of the wave equation with space complexity (qubit number) $O(d\log(l/a))$.
The time complexity (gate number) of the algorithm scales as $\tilde{O}(td^2/a)$, where the time evolution is based on the Hamiltonian simulation algorithm~\cite{Berry2015a} and the $\tilde{O}$ notation indicates the suppression of logarithmic factors.
In the work by Suau et al.~\cite{suau2021practical}, the authors provide a resource estimation for a practical implementation of~\cite{costa2019quantum} and obtain prohibitively large prefactors in the asymptotic gate complexity.

In this article, we revisit the quantum solution to the wave equation given in~\cite{costa2019quantum} with the intent of providing a robust benchmark for evaluating the performance of quantum architectures as the industry continues to advance. 
We focus on the one-dimensional, isotropic, acoustic wave equation as a PDE of relevance to both industry and the scientific community.
We provide a quantum algorithm that is analogous to the classical spectral method.
We use efficient approximations to derive, implement and benchmark quantum circuits for simulating the isotropic wave equation with periodic boundary conditions.
The simplicity of this PDE makes it ideally suited to gauge the viability of quantum computing hardware in simulating the wave equation using the spectral method beyond classical memory limits.

This article has the following structure.
Section~\ref{section: Methods} contains a derivation of the quantum circuits used to simulate the wave equation, including the state preparation of the initial conditions in Sec.~\ref{section: state prep}, and approximation of the time evolution operator in Sec.~\ref{section: time evolution}.
In Sec.~\ref{section: Results} we present the results from running these quantum circuits, using both a depolarizing noise model and real hardware experiments.
Finally in Sec.~\ref{section: Discussion} we discuss further improvements to our approach through optimization of the state preparation and quantum Fourier transform circuits.

\begin{figure*}
\centering
\includegraphics[width=165.315mm]{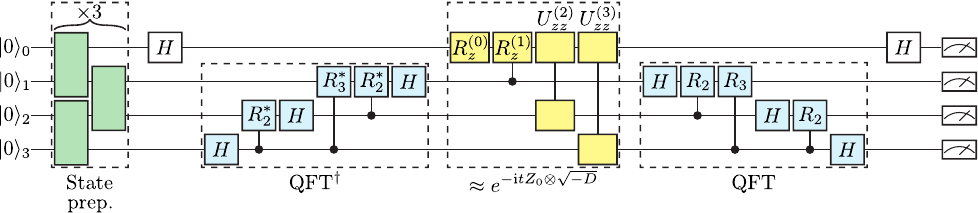}
\caption{\label{fig: fourier circuit}
Circuit for approximately simulating the wave equation~\eqref{eq: discretized wave equation} on a quantum computer for $n+1 = 4$ qubits.
The circuit can be split into three parts:
initial state preparation, time evolution and final state readout via qubit measurements.
We define the initial state in terms of a variational brickwall circuit of depth $\text{floor}\left(\log_{2}(n+1)\right) + 1$, and is composed of generic nearest-neighbor two-qubit unitaries.
The time evolution consists of a diagonal operator approximating $\exp(-\text{i}t Z_{0} \otimes \sqrt{-D})$ preceded by the inverse quantum Fourier transform QFT$^{\dag}$ and proceeded by the QFT.
Here $H$ is the Hadamard gate, $R_{\kappa} = \text{diag}(1, \exp(\text{i} 2 \pi / 2^{\kappa}))$, $R_{\kappa}^{*}$ is the conjugate of $R_{\kappa}$, $R_{z}^{(\kappa)} = \exp(-\text{i} \theta_{\kappa} Z)$, $U_{zz}^{(\kappa)} = \exp(-\text{i} \theta_{\kappa} Z \otimes Z)$, $\theta_{0} = 3 \pi t$, $\theta_{1} = -8 \pi t$, $\theta_{2} = -2 \pi t$ and $\theta_{3} = -\pi t$.
These gates are derived from Eq.~\eqref{eq: approximation operator new}.
Note that we label the qubits $0$, $1$, \ldots, $n$ from top to bottom.
}
\end{figure*}

\section{Methods}\label{section: Methods}

In this section, we start by mapping the wave equation onto a Schr\"{o}dinger equation, taking advantage of the eigendecomposition of the discrete Laplacian.
We then describe the state preparation in Sec.~\ref{section: state prep} that uses a logarithmic depth ansatz with a brickwall structure.
In Sec.~\ref{section: time evolution} we are able to use the small-angle approximation in order to derive efficient circuits for the time evolution operator.

We discretize space into a one-dimensional grid of length $l$ and lattice spacing $a$, with periodic boundary conditions and $N=l/a$ grid points in each dimension.
Note that the restriction to periodic boundary conditions does not compromise the utility of this approach in practical applications.
Since the number of grid points scales exponentially with the number of qubits, it is not unrealistic to make the grid containing the solution larger, such that the wavefield does not reach the boundary in the time frame of interest.

We write the discretized one-dimensional (1D) isotropic acoustic wave equation~\eqref{eq: acoustic wave equation} with $c = 1$ as
\begin{equation}\label{eq: discretized wave equation}
 \frac{\partial^{2}}{\partial t^{2}} \psi(x_j,t) = \Delta_a\psi(x_j,t),
\end{equation}
where $\psi(x_j,t)$ is the discretized wavefield at position $x_j$ and time $t$, and $\Delta_a$ is a discretized Laplacian operator which depends on the lattice spacing $a$.
After specifying the initial conditions for $\psi(x_{j},t)$ and velocity $\partial_t\psi(x_{j},t)$ at $t=0$, our goal is to compute $\psi(x_{j},t)$ at $t>0$.
For ease of notation, we drop the dependency on $x_j$ and $t$ within the argument of the wavefield and velocity.
Following \cite{costa2019quantum}, we construct a Schr\"odinger equation
\begin{equation}\label{eq: schroedinger for wave equation}
 \frac{\partial}{\partial t}
 \begin{pmatrix}
 \psi\\
 \phi
 \end{pmatrix}
 =
 -\text{i} \begin{pmatrix}
 0 & \sqrt{-\Delta_{x}}\\
 \sqrt{-\Delta_{x}} & 0
 \end{pmatrix}
 \begin{pmatrix}
 \psi\\
 \phi
 \end{pmatrix},
\end{equation}
where we introduce a new variable
$\phi = \text{i}(-\Delta_a)^{-1/2}\partial_t\psi$
and identify the wave-equation Hamiltonian $H_{\text{W}}$ with the Hermitian operator
\begin{equation}
 H_{\text{W}} = \begin{pmatrix}
 0 & \sqrt{-\Delta_{a}}\\
 \sqrt{-\Delta_{a}} & 0
 \end{pmatrix}.
\end{equation}
We associate the fields $\psi$ and $\phi$ with a normalized quantum state $\ket{\Phi} \propto (\psi, \phi)^T$, where $\ket{\Phi}$ is the vector containing the values of $\Phi$ on all spatial grid points and $(\cdots)^T$ denotes the transpose.
Note that for periodic boundary conditions and a second order finite difference approximation, the Quantum Fourier Transform (QFT)~\cite{ChuangQuantum2010} diagonalizes the Laplacian (see Appendix~\ref{app:Diagonalization}) as
\begin{equation}\label{eq: diagonalised laplacian}
    \Delta_a = \text{QFT} \, D \, \text{QFT}^{\dagger},
\end{equation}
where $D$ is a diagonal operator of eigenvalues $E_{k}$ indexed by the wavenumbers $k$:
\begin{equation} \label{eq: eigenvalues}
 E_{k} = -4 N^{2} \sin^{2}\left(\frac{\pi k}{N}\right).
\end{equation}
We express the Hamiltonian $H_{\text{W}}$ in terms of $D$ as 
\begin{align}
    H_{\text{W}} &= X \otimes \sqrt{-\Delta_a} \nonumber \\
    &=  \left( H \otimes \text{QFT} \right) \left( Z \otimes \sqrt{-D} \right) \left( H \otimes \text{QFT}^{\dagger} \right),
\end{align}
where $X, Z$ are Pauli matrices, $\otimes$ is the Kronecker product, $H$ is a Hadamard operator and $D$ contains the eigenvalues in Eq.~\eqref{eq: eigenvalues}.
By solving the Schr\"odinger equation, the state of the system after a time $t$ can be expressed as 
\begin{align}\label{eq: wave equation solution}
 \ket{\Phi(t)} &= \left( H \otimes \text{QFT} \right) e^{-\text{i}t Z \otimes \sqrt{-D}} \left( H \otimes \text{QFT}^{\dagger} \right) \ket{\Phi(0)}
\end{align}
where $\ket{\Phi(0)}$ encodes the initial function, see Sec.~\ref{section: state prep} for more details.
A quantum circuit for approximately simulating the wave equation is shown in Fig.~\ref{fig: fourier circuit}.
The circuit involves a state preparation circuit to prepare $\ket{\Phi(0)}$, followed by a circuit, discussed in Sec. \ref{section: time evolution}, that implements the time evolution.
Finally, there is a measurement procedure to retrieve information about the state $\ket{\Phi(t)}$ that contains the wavefield. 

This approach requires $n+1$ qubits to represent the wave equation solution on $N = 2^{n}$ spatial grid points.
Measuring the top qubit of Fig.~\ref{fig: fourier circuit} in state $\ket{0}$ or $\ket{1}$ determines the state $\ket{\psi}$ or $\ket{\phi}$ encoded into the other qubits respectively.
The number of qubits required to represent $\psi$ and $\phi$ as a quantum state scales as $O(\log(l/a))$.
The circuit complexity of the state preparation circuit is highly dependent on the details of the initial conditions, e.g., the smoothness of $\psi(x_j,0)$ and $\phi(x_j, 0)$.
In the general case for finite $\phi(x_j, 0) = \text{i}(-\Delta_{a})^{-1/2} \partial_{t} \psi(x_{j}, 0)$, a quantum linear systems algorithm is required for state preparation~\cite{costa2019quantum}.
Additionally, obtaining information about $\partial_t \psi$ using the form of $\phi$ given above requires additional post-processing~\cite{costa2019quantum}.

\subsection{State preparation}\label{section: state prep}

The state preparation circuit maps the product state $\ket{0}^{\otimes n+1}$ to the normalized state $\ket{\Phi(x_j,0)} \propto (\psi(x_j,0),\phi(x_j,0))^T$.
Here, we assume $\psi(x,0)$ is a Ricker wavelet~\cite{ricker1953form} and w.l.o.g. set the initial velocity $\phi(x_j,0)=0\;\forall j$.
This waveform is commonly used as a benchmark for wave equation solvers~\cite{wang2015generalized} and has the form
\begin{equation}\label{eq: ricker wave form}
    \psi(x,0) = \frac{2}{\sqrt{3 \Sigma}\pi^{\frac{1}{4}}} \left(1 - \left(\frac{x - \mu}{\Sigma}\right)^2 \right)
    e^{-\frac{(x - \mu)^2}{2\Sigma^2}}.
\end{equation}
We set $\mu = 0.5$, $\Sigma = 0.1$ and $x \ \in \ [0,1)$ and use a parameterized quantum circuit (PQC) with a brickwall structure to prepare the state. 
For further details on the PQC used for state preparation, see Appendix~\ref{app:PQC}.

\begin{figure*}[t]
    \centering
    \includegraphics[width = 0.8\textwidth]{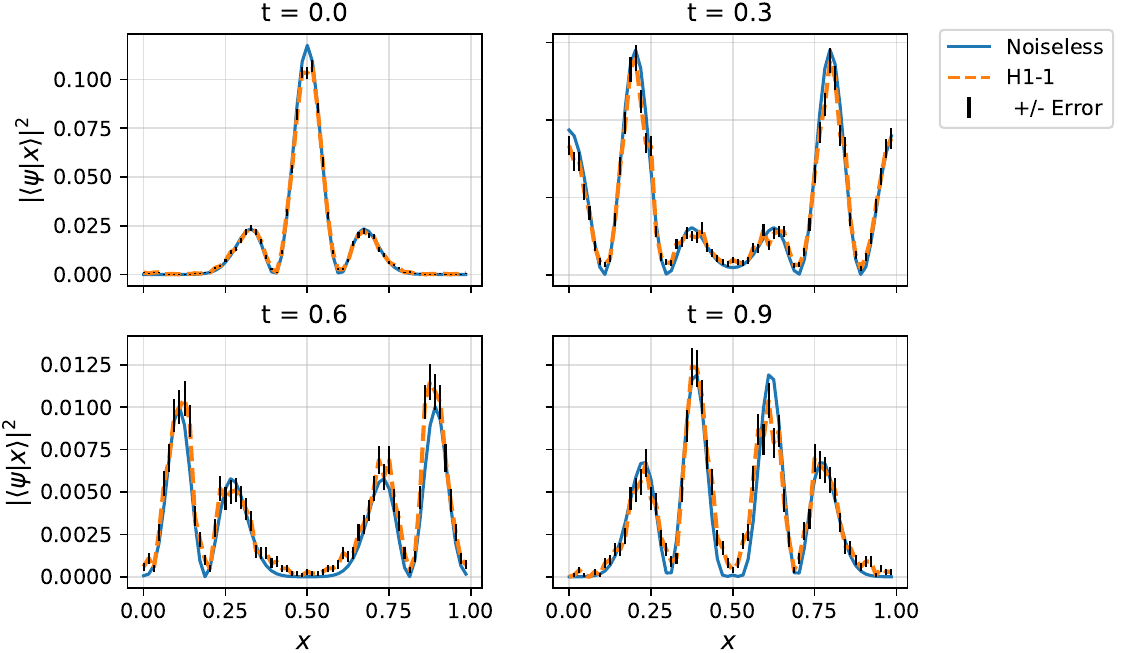}
    \caption{\label{fig: quantum solution to wave equation}
    Quantinuum system model H1-1 hardware results for simulating a Ricker wavelet according to the wave equation at different times $t$ and $n=6$ qubits.
    The exact solution (blue solid line) is shown alongside the approximate solution (orange dashed line) which is sampled from H1-1 with $N_{\text{shots}} = 10,000$ shots.
    Samples are collected by measuring all qubits.
    The black error bars represent $\pm \epsilon_{\text{MC}}$, where $\epsilon_{\text{MC}} = \sigma / \sqrt{N_{\text{shots}}}$ is the Monte Carlo sampling error with $\sigma$ being the standard deviation.
    Circuits were evaluated on the Quantinuum system model H1-1~\cite{H1datasheet} on December 1st-6th, 2023.
    All shots are included in the figure, however we only display the un-normalized state corresponding to the wavefield $\psi(x, t)$.
    Additional details about the experiments can be found in Appendix~\ref{app:experiment details}.
    }
\end{figure*}

\subsection{Time evolution}\label{section: time evolution}

The main challenge of implementing the time evolution step is the realization of $\exp(-\text{i}t Z \otimes \sqrt{-D})$ as a quantum circuit with sufficiently shallow depth to run on NISQ devices.
Note that in the following, we use the qubit labels explained in the figure caption of Fig.~1 as subscripts whenever required to clarify which qubits states correspond to and operators act on.
By expanding in the Fourier basis labeled by the wavenumber $k$, we write the diagonal operator as
\begin{equation}\label{eq: diagonal operator}
    e^{-\text{i}t Z_{0} \otimes \sqrt{-D}} = \sum_{k = -N/2}^{N/2 - 1} e^{-\text{i}t2N \sin(\pi k / N) Z_{0}} \otimes \ket{k}\bra{k},
\end{equation}
where $\ket{k}\bra{k}$ acts on qubits $1$ to $n$, see Fig.~\ref{fig: fourier circuit}.

For sufficiently smooth initial states expanded in the Fourier basis, the amplitudes corresponding to small wavenumbers dominate those of large wavenumbers.
Therefore, we can make the small-angle approximation $\sin(\pi k / N)\approx \pi k/N$ without significant loss of accuracy and approximate the diagonal operator as
\begin{equation}\label{eq: approximation}
    e^{-\text{i}t Z_{0} \otimes \sqrt{-D}} \approx \sum_{k = -N/2}^{N/2 - 1} e^{-\text{i}t 2 k \pi Z_{0} } \otimes \ket{k}\bra{k}.
\end{equation}
For the representation of $\ket{k}$, we define the state for the first qubit of $\ket{k}$ (qubit $1$ in Fig.~\ref{fig: fourier circuit}) to determine the sign of $k$, i.e. $\ket{0}_{1}$ for $+\textit{ve}$ and $\ket{1}_{1}$ for $-\textit{ve}$.
The other qubits of $\ket{k}$ (qubits $2$ to $n$ in Fig.~\ref{fig: fourier circuit}) determine the value of $l \in \{0, 1, \ldots, N/2-1\}$ using binary representation $\ket{l} = \ket{b_{2}, b_{3}, \ldots, b_{n}}$, where binary variable $b_{q} \in [0,1]$ and,
\begin{equation}\label{eq: binary rep}
    l = \sum_{q = 2}^{n} b_{q} 2^{n-q}.
\end{equation}
To take negative wavenumbers into account, we apply a phase factor to Eq.~\eqref{eq: approximation} when qubit $1$ is in state $\ket{1}$,
\begin{align}\label{eq: shifted diagonal operator}
    &\sum_{k = -N/2}^{N/2 - 1} e^{-\text{i}t 2 k \pi Z_{0} } \otimes \ket{k}\bra{k} \nonumber \\
    &= \! \! \sum_{l = 0}^{N/2 - 1} e^{-\text{i}t 2 l \pi Z_{0}} \left( \ket{0}_{1}\bra{0}_1 + e^{\text{i}t N \pi Z_{0}}\ket{1}_{1}\bra{1}_{1} \right) \otimes \ket{l}\bra{l}
\end{align}
where $\ket{l}\bra{l}$ acts on qubits $2$ to $n$.
To realize the $l$-dependent phases, we insert Eq.~\ref{eq: binary rep} in Eq.~\ref{eq: shifted diagonal operator} and replace $b_{q}$ by its operator representation: $(\mathds{1}_{q}-Z_{q})/2$.
This leads to our final expression for the time evolution operator,
\begin{align}\label{eq: approximation operator new}
    &\sum_{k = -N/2}^{N/2 - 1} e^{-\text{i}t 2 k \pi Z_{0} } \otimes \ket{k}\bra{k} \nonumber \\
    &= \! \! e^{-\text{i}t (2^{n-1}-1) \pi Z_{0}} \left(\ket{0}_{1}\bra{0}_{1} + e^{\text{i}t N \pi Z_{0}}\ket{1}_{1}\bra{1}_{1}\right) \prod_{q=2}^{n} e^{\text{i}t 2^{n-q} \pi Z_{0} Z_{q}}
\end{align}
where we have used $\sum_{q=2}^{n} 2^{n-q} = 2^{n-1}-1$.
Equation~\eqref{eq: approximation operator new} gives an approximation that contains a product of $n$ two-qubit gates and is accurate for $k/N \ll 1$.
The two-qubit gate count of the entire time evolution operator is dominated by the QFT that scales as $O(n^2)$.

\section{Results}\label{section: Results}

\begin{figure}
     \centering
     \includegraphics[width =\linewidth]{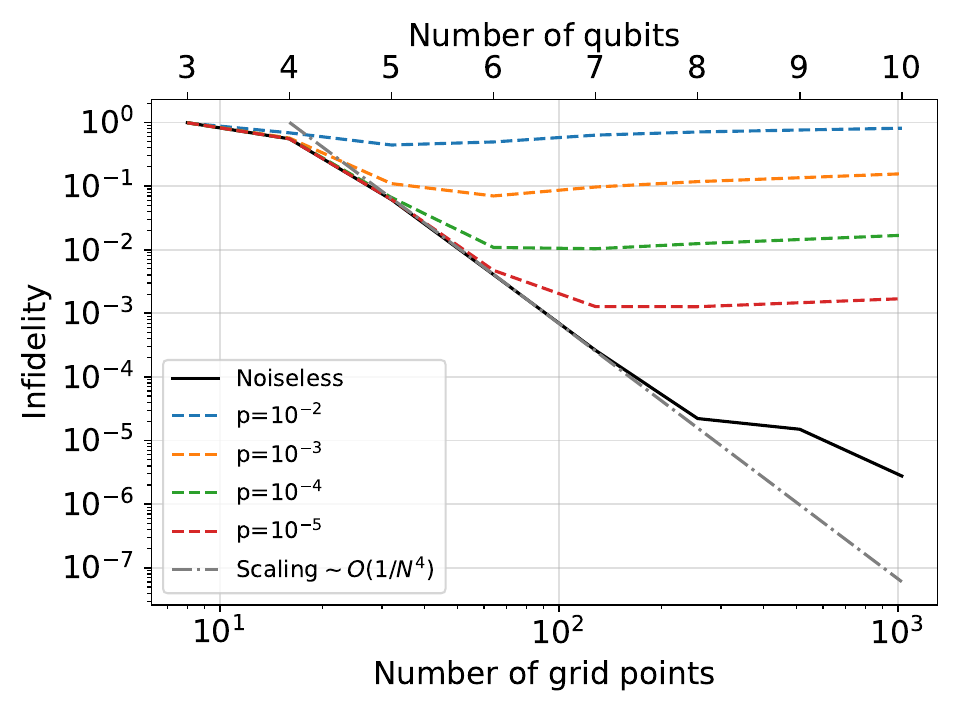}
     \caption{\label{fig: error vs N}
     Infidelity~\eqref{eq: state error} as a function of the number of grid points $N$ and number of qubits $n$ at $t=1$.
     The noiseless approximate circuits are evaluated using a statevector simulator (solid black line).
     In the noiseless case, the infidelity reduces exponentially with the number of qubits $n$.
     The PQC is used for state preparation in all cases. 
     A density matrix simulator is used to sample noisy approximate circuits with a depolarizing noise model with depolarizing probabilities $p$ (dashed lines).
     We observe a minimum in the noisy case when the gain in approximation accuracy cancels out with the loss in error from gate noise.
     We also plot the asymptotic scaling of infidelity with respect to $N$ for fixed $t$ given by $\epsilon \sim O(1/N^4)$ (dashed-dotted) -- a full derivation of this scaling is given in Appendix~\ref{app:IFD}.
     We observe good agreement between noiseless simulations and asymptotic scaling.
     Deviations in the small $N$ regime are due to inaccurate small angle approximations whilst deviations in the large $N$ regime are due to the \textit{approximate} state preparation performed by the PQC.
     Further details about the PQC state preparation performance can be found in Appendix~\ref{app:PQC}.}
\end{figure}

\begin{figure}
     \centering
     \includegraphics[width=\linewidth]{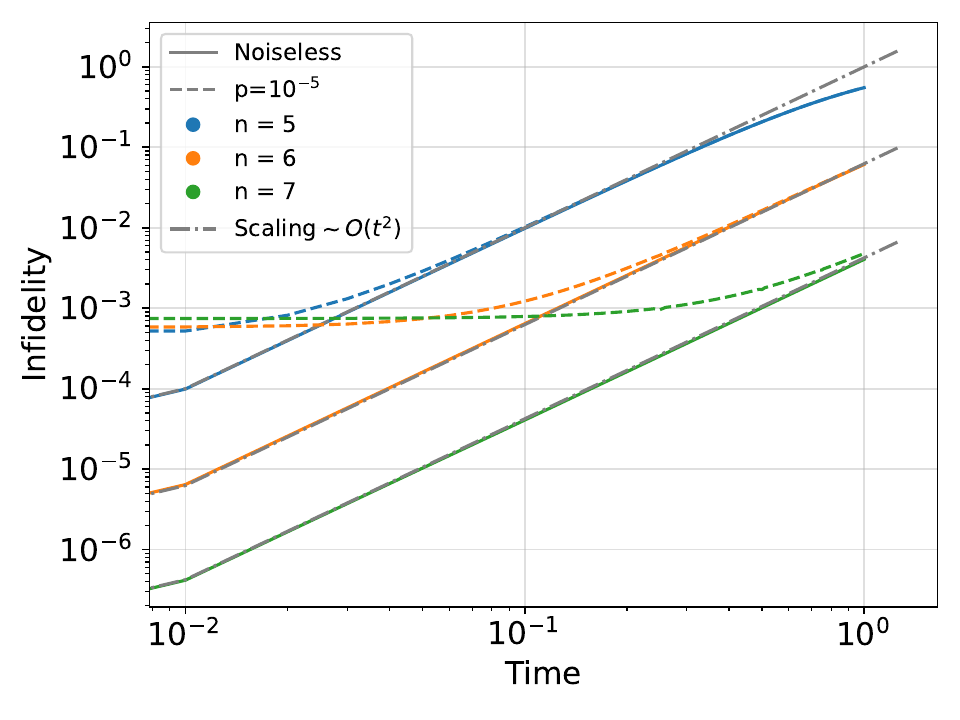}
     \caption{\label{fig: error vs t}
     Infidelity~\eqref{eq: state error} as a function of time $t \ \in \ [0, 1]$ with time step $\delta = 0.01$ for noiseless (solid lines) and noisy circuit simulations (dashed lines).
     In the noiseless case (solid lines), the infidelity scales as $O(t^2)$ with a prefactor determined by the number of qubits $n$.
     A density matrix simulator is used to sample from noisy approximate circuits (dashed lines) using a depolarizing noise model with depolarizing probability $p=10^{-5}$.
     We also plot the scaling of infidelity with respect to time for fixed numbers of grid points given by $\epsilon(t) \sim O(t^2)$ (dashed-dotted) -- a full derivation of this scaling is given in Appendix~\ref{app:IFD}.
     The slight deviation for $n=5$ is due to low number of grid points, leading to poor approximation performance for $t \approx 1$.}
\end{figure}

In this section, we present the solution to the wave equation obtained using the quantum algorithm described in Sec.~\ref{section: Methods} evaluated using numerically exact statevector simulations and real hardware.

The quantum circuit for the procedure, given in Fig.~\ref{fig: fourier circuit}, can be summarized as follows: a Ricker wavelet is initialized using a shallow depth PQC that has been classically optimized using the method described in Appendix~\ref{app:PQC}; a transformation to the Fourier basis is performed using an exact inverse QFT; the diagonal operator is applied either exactly or approximately; the basis is transformed back into real space by a QFT, and finally the state is read out by measuring the qubits.

Figure~\ref{fig: quantum solution to wave equation} shows the quantum solution to the wave equation given by the square absolute value of amplitudes $\lvert\psi(x_{j},t)\rvert^2$, i.e. the probabilities of sampling the bitstrings associated to positions $x_{j}$.
Circuits are implemented on the Quantinuum system model H1-1 computer which has single- and two-qubit gate infidelities $2.95 \times 10^{-5}$ and $1.39 \times 10^{-3}$, respectively --- as of July 2023~\cite{H1datasheet}.
Additional details including resource count and shot times can be found in Appendix~\ref{app:experiment details}.
Although the whole state is measured, only the wavefield i.e. $\psi$ is shown.
The error bars (black solid lines) represent $\pm \epsilon_{\text{MC}}$, where $\epsilon_{\text{MC}} = \sigma/\sqrt{N_{\text{shots}}}$ is the Monte Carlo sampling error and $\sigma$ is the standard deviation of the expectation values for each bitstring~\cite{NumericalRecipes}.
We find good agreement between the exact noiseless results and those sampled from the device, with only a small degradation in solution quality over time. 
The smaller error bars at $t=0$ are due to the absence of the time evolution operator.

To quantify the error produced by approximating the diagonal operator, we define the infidelity as,
\begin{equation}\label{eq: state error}
 \epsilon(t) =  1 - \bra{\Phi(t)}\rho(t)\ket{\Phi(t)},
\end{equation}
where $\ket{\Phi(t)}$ is the result of evolving the state using the exact diagonal operator, and $\rho$ is the mixed state prepared using the approximate diagonal operator.
In the noiseless case, $\rho(t) = \ket{\Psi(t)}\bra{\Psi(t)}$ where $\ket{\Psi(t)}$ is prepared according to Eq.~\eqref{eq: wave equation solution} using the approximate diagonal operator.
Equation~\eqref{eq: state error} depends on the entirety of $\ket{\Phi(t)}$ and $\rho(t)$ i.e. both $\psi$ and $\phi$.

Figure~\ref{fig: error vs N} shows the infidelity given in Eq.~\ref{eq: state error} as a function of $N$ at $t=1$.
As $N$ increases, we expect the small-angle approximation $k/N \ll 1$ to become more accurate.
This behaviour is clearly observed in the noiseless case (black solid line), where the infidelity reduces exponentially with $n$.
To investigate the influence of noisy gates in our circuits, we introduce a depolarizing channel~\cite{ChuangQuantum2010} with probability $p$ on all two-qubit gates within the circuit, including those involved in the state preparation.
The resulting infidelities are shown in Fig.~\ref{fig: error vs N} (dashed lines). 
We observe an interplay between decreasing errors with more qubits, and increasing errors due to higher circuit depth.
At each rate of depolarizing noise $p$, we find an optimal qubit number at which the infidelity is at its minimum.
These results highlight the intricate trade-off between decreasing the infidelity and exposure to gate noise in the NISQ regime, a behavior previously analyzed in \cite{KiEtAl23}.

Figure~\ref{fig: error vs t} shows the infidelity as a function of $t$ for $n\in\{5,6,7\}$, with time step $\delta t = 10^{-2}$ and depolarizing probability $p=10^{-5}$.
At earlier times, the depolarizing noise is the dominant contribution to the observed infidelity.
At larger times, the noisy results align with the noiseless results and we observe that $\epsilon(t) \propto t^2$, see Appendix~\ref{app:IFD} for a full derivation of this scaling.
In both noiseless and noisy cases, increasing $N$ (or equivalently $n$) leads to a smaller infidelity for a fixed time $t$.  

\section{Discussion}\label{section: Discussion}

In this work, we present quantum circuits for simulating the one-dimensional isotropic acoustic wave equation which are suitable for NISQ devices.
We utilize physically motivated approximations that become more accurate as the number of qubits increases.

We efficiently prepare the initial state by a variational optimization of a logarithmic depth PQC.
One approach to potentially circumvent barren plateaus as $N$ increases is to use an iterative approach, such as multigrid renormalization~\cite{Lubasch2018}. 
By starting with a trainable PQC, optimizing PQCs with increasing $N$ could be achieved by initializing with the previously optimized (and trainable) PQC representing a coarser grid.
Alternatively, quantum signal processing could be used for state preparation~\cite{mcardle2022quantum}.
Although the circuit structure is less flexible and typically has a higher depth than in the PQC approach, quantum signal processing can prepare the state with less error and does not suffer from barren plateaus --- making it more suitable in the fault tolerant regime.

By using circuits that approximate the QFT \cite{hales2000improved}, we can reduce the cost of the time evolution operator further.
Keeping only the $O[\log(n/\epsilon)]$ largest controlled angle rotations connected to each qubit produces a spectral norm error of $O(\epsilon)$ \cite{nam2020approximate}.
Also, we can take advantage of the H-series hardware, and replace the conditional gates found in the final QFT (shown in Fig.~\ref{fig: fourier circuit}) with measurements and classically controlled single-qubit gates --- a procedure known as semi-classical QFT \cite{griffiths1996semiclassical, chiaverini2005implementation}.
This removes all two-qubit gates from the QFT, and instead applies a phase if and only if the conditional qubit is measured in $\ket{1}$.
Since the measurement collapses the state, this procedure is only applicable if the outcome is measured immediately afterwards, i.e. when converting back from Fourier to real space.
Both of these approaches for optimizing the QFT circuits could be applied simultaneously to bring about further reductions in gate count.

Since an analytic solution to the PDE of this form exists, these circuits serve as an ideal benchmark to test the viability of quantum computers for simulating PDEs by comparing their output with the known solution. 
Additionally, our algorithm would be appropriate as an application in the framework of application-oriented performance benchmarking \cite{lubinski2023applicationoriented}, where the performance of quantum computing hardware for various system sizes, evolution times and different initial waveforms could be tested.
Although we focus on one-dimensional grids, expanding to higher-dimensional grids is of notable interest.
However, because the square-root of the Laplacian is not separable in higher dimensions, further approximations can be made to realize separability.
One potential avenue is to utilize the ``alpha max plus beta min'' algorithm~\cite{abdulsatar2021asic} which makes an accurate approximation to the square-root of the higher-dimensional Laplacian via a linear combination of the arguments.
Once this separation has been achieved, we can implement the smoothness approximation in each dimension.
We hope that the work given in this paper bridges the gap between the theoretical and practical uses of quantum devices, while exploring hardware implementations of PDE solvers.

\begin{figure*}
\centering
\includegraphics[width=183.02mm]{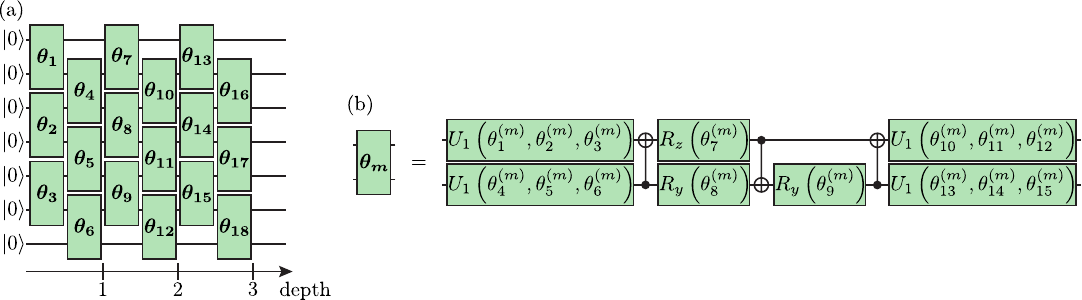}
    \caption{\label{fig: pqc architecture}
PQC for 7 qubits as considered in the experiments.
(a) Brickwall circuit of depth 3 composed of 18 two-qubit gates with variational parameters $\boldsymbol{\theta_{1}}, \boldsymbol{\theta_{2}}, \ldots, \boldsymbol{\theta_{18}}$.
(b) Each two-qubit gate consists of generic single-qubit gates $U_{1}$, each parameterized by 3 variational angles (excluding the physically irrelevant global phase)~\cite{ChuangQuantum2010}, CNOTs, and single-qubit rotation gates $R_{z}(\theta) = \exp(-\text{i} \theta Z)$ and $R_{y}(\theta) = \exp(-\text{i} \theta Y)$.
}
\end{figure*}

\acknowledgements
We are grateful to Marcello Benedetti, Eric Brunner and Yuta Kikuchi for valuable feedback on the article.

\appendix
\section{Diagonalization of the discretized Laplacian}
\label{app:Diagonalization}

We derive the discretized Laplacian $\Delta_a$ using the central difference approximation
\begin{equation}\label{eq: discretized laplacian}
    [\Delta_a \psi]_j = \frac{\psi_{j-1} - 2\psi_j + \psi_{j+1}}{a^2},
\end{equation}
which has truncation error $O(a^2)$ and becomes equivalent to the second derivative  with respect to continuous space in the limit $a \rightarrow 0$.
It can be shown that Eq.~\eqref{eq: discretized laplacian} with periodic boundary conditions $j+N=j$ is diagonal in the basis of discretized plane waves,
\begin{equation} \label{eq: discretised plane waves}
    f(x_j) = \frac{1}{\sqrt{N}} e^{\text{i} \frac{2 \pi}{N} k j},
\end{equation}
where $N = 1 / a$ and integer wavenumbers $k \in \{-N/2, -N/2+1, \ldots, N/2-1\}$).
For each wavenumber $k$, the associated eigenvalue is $E_k = -4N^2 \sin^2(\pi k / N)$.
Since Eq.~\eqref{eq: discretised plane waves} for different $k$ is equivalent to the columns of the Quantum Fourier Transform (QFT), then we can decompose Eq.~\eqref{eq: discretized laplacian} as follows,
\begin{equation}\
    \Delta_a = \text{QFT} \, D \, \text{QFT}^{\dagger},
\end{equation}
where $D$ is a diagonal operator containing the eigenvalues $E_k$.

\section{Parameterized Quantum Circuits for state preparation}\label{app:PQC}

Parameterized Quantum Circuits (PQCs) allow shallow but accurate representations of otherwise deep quantum circuits to be found variationally~\cite{Cerezo2021}.
Given a circuit with adjustable angles $\boldsymbol{\theta}$, a PQC prepares $U(\boldsymbol{\theta})|0\rangle^{\otimes n} = |g(x)\rangle$ up to error $\gamma$ for function $g(x)$.
This is typically done by classical updates to $\boldsymbol{\theta}$ with respect to a cost function, $C(\boldsymbol{\theta})$ ,that is usually evaluated with the help of a quantum computer.
Due to the relatively small number of qubits used in the experiments of this article ($n \le 10$), we update the PQC classically using a statevector simulation. 

The success of PQCs is highly dependent on a variety of factors, including the chosen structure, cost function, optimizer and values of the initial angles $\boldsymbol{\theta}_\text{init}$.
We use a brickwall ansatz with logarithmic depth in the number of qubits for our PQC as defined in Fig.~\ref{fig: fourier circuit}.
Figure~\ref{fig: pqc architecture} shows the specific PQC architecture used in the experiments.
The cost function,
\begin{equation}
    C(\boldsymbol{\theta}) = 1 - \text{Re}\left( \langle g(x)|U(\boldsymbol{\theta})|0\rangle^{\otimes n}\right),
\end{equation}
is updated via a pseudo-global Newton optimizer, the Limited-memory Broyden–Fletcher–Goldfarb–Shanno (L-BFGS) algorithm \cite{liu1989limited}, with 100,000 optimization steps.
Initial values $\boldsymbol{\theta}_{\text{init}}$ are sampled from a uniform distribution between $[0,1)$.
For larger numbers of qubits we recommend a more sophisticated initialization scheme for $\boldsymbol{\theta}_\text{init}$ to make the optimization problem easier and avoid barren plateaus --- a phenomenon whereby the cost function landscape becomes exponentially flat~\cite{McClean2018}.

\begin{figure}[h]
     \centering
     \includegraphics[width=\linewidth]{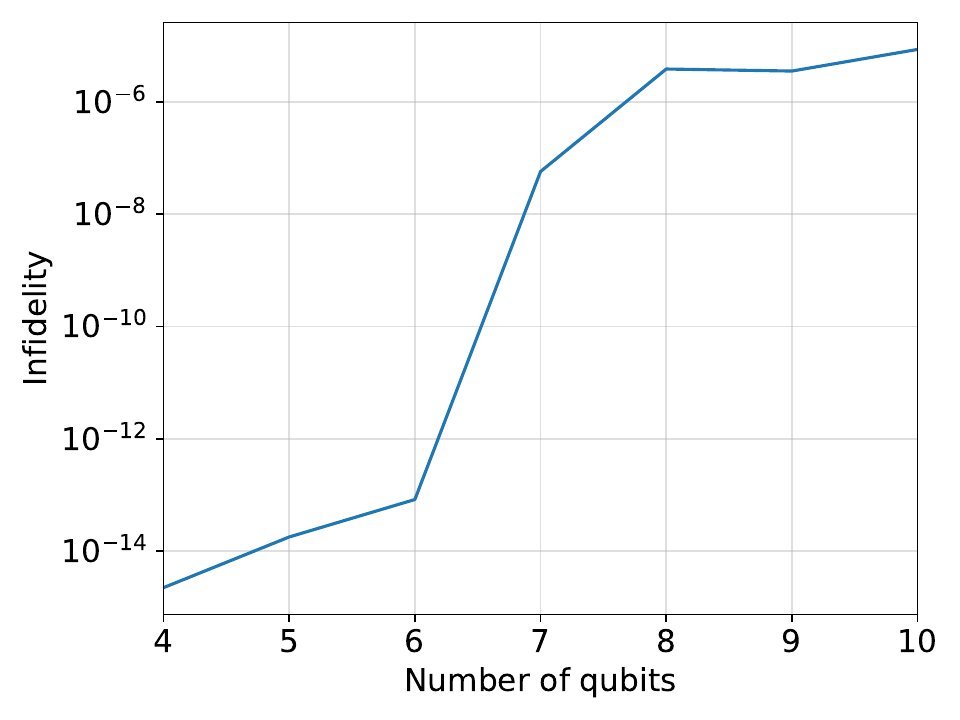}
     \caption{\label{fig: pqc infidelity}
     Infidelity~\eqref{eq: state error} between the exact state and the state prepared by a PQC as a function of the number of qubits.
     The PQC ansatz contains a brickwall circuit with logarithmic depth in number of qubits and is optimized using a pseudo-global Newton optimizer with a fixed number of optimization steps, as explained in Appendix B.
     }
\end{figure}

Figure~\ref{fig: pqc infidelity} shows the infidelity between the exact state and the state prepared by the PQC.
We can see a higher infidelity as the number of qubits increase.
This is because we fix the number of steps, whilst the optimization requires a larger amount of steps in order to reach to same performance across higher numbers of qubits.
The performance scaling of this optimization procedure with respect to number of qubits is beyond the scope of this article, but we suspect that alternative methods of state preparation, such as quantum signal processing, would perform better for higher numbers of qubits. 

\section{Experiment details}\label{app:experiment details}

\begin{table}[h!]
\small
\caption{\label{tab: nisq costs}
Resource costs and shot time estimates for circuits of different evolution times $t$ run on Quantinuum's H1-1 quantum computer.
Circuits were compiled in the series H1-1 gateset, \{RZ, PhasedX, RZZ\}, using Pytket with optimization level 2.}
\begin{ruledtabular}
\begin{tabular}{@{}lccc@{}}
evolution time & no.~gates & no.~2Q gates & shot time (s) \\
\cmidrule(){1-4}
$t = 0$ & $176$ & $54$ & $0.2708$ \\
$t \in \{0.3,0.6,0.9\}$ & 211 & 71 & 0.5255 \\
\end{tabular}
\end{ruledtabular}
\end{table}

In this section, we provide further details on the parameters used in the experiments performed on Quantinuum's series H1-1 device as shown in Fig.~\ref{fig: quantum solution to wave equation}. 
Table~\ref{tab: nisq costs} shows the resource costs and shot times required to implement the $n=6$ qubit circuits at various time steps.
At $t=0$ the circuit simply consists of the state preparation PQC whereas the circuits at $t=0.3, 0.6, 0.9$ also contain the QFT (and inverse) in addition to the approximate diagonal operator.
Note that all of the finite time $(t>0)$ circuits have the same circuit depth and gate count.

\begin{figure}[h]
     \centering
     \includegraphics[width=\linewidth]{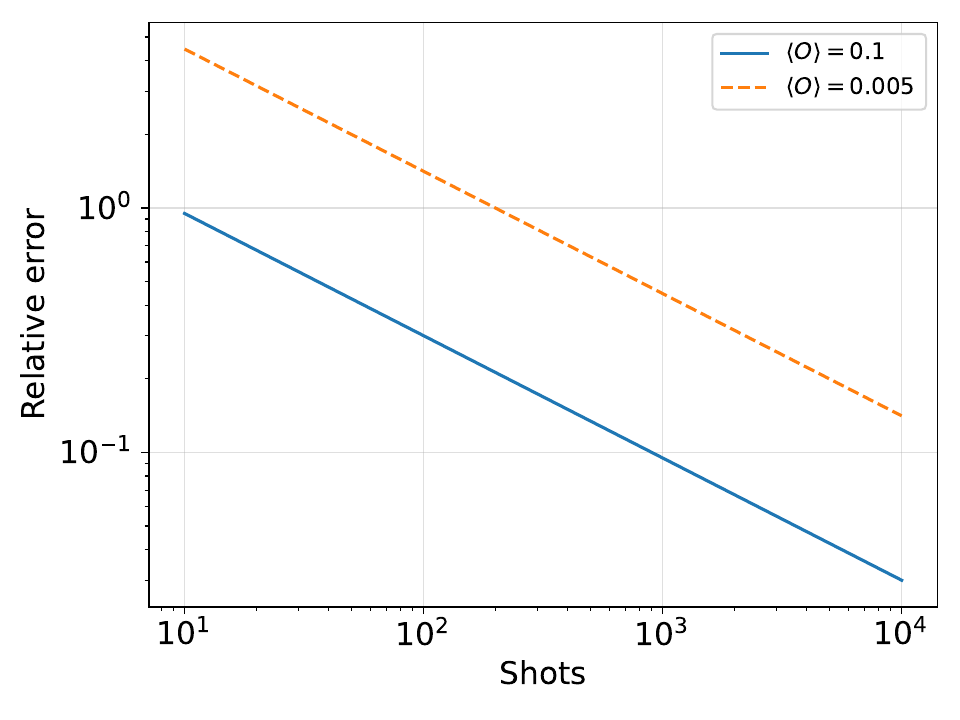}
     \caption{\label{fig: rel error vs samples}
     Relative Monte Carlo sampling error~\ref{eq:rel Monte Carlo error} for the largest ($\langle O \rangle = 0.1$) and smallest ($\langle O \rangle = 0.005$) peaks from Fig.~\ref{fig: quantum solution to wave equation} with varying numbers of shots.}
\end{figure}

The time per shot is computed from averaging 500 shots ($t=0$) and 200 shots ($t=0.3,0.6,0.9$). 
Although 10,000 shots were used to produce Fig.~\ref{fig: quantum solution to wave equation} at each time step, a smaller number of shots could also be used to reproduce similar results.
This is shown in Fig.~\ref{fig: rel error vs samples} whereby we compute the relative Monte Carlo sampling error, 
\begin{equation}\label{eq:rel Monte Carlo error}
    \epsilon_{\text{rel}} = \frac{\sigma}{\langle O \rangle \sqrt{N_{\text{shots}}}},
\end{equation}
with an increasing number of shots for the largest ($\langle O \rangle = 0.1$) and the smallest ($\langle O \rangle = 0.005$) peaks from Fig.~\ref{fig: quantum solution to wave equation}.
These peaks were chosen since they give a bound on the amount of samples required to resolve the smallest and largest amplitude features.
We can see that to observe $\epsilon_{\text{rel}} \sim 0.1$ ($10\%$) for $\langle O \rangle = 0.1$ requires $\sim 1000$ shots whilst $\langle O \rangle = 0.005$ requires an order of magnitude more shots to observe the same relative error.

\section{Asymptotic scaling of infidelity}\label{app:IFD}

The initial state prepared by the PQC can be expressed in the Fourier basis as
\begin{equation}\label{eq: init state}
    \ket{\psi_{\text{init}}} = \frac{1}{\sqrt{2}} \sum_k \left(c_{0,k}\ket{0} + c_{1,k}\ket{1}\right) \otimes \ket{k},
\end{equation}
where $c_{i,k}$ are the real space amplitudes of the initial function multiplied by the Fourier phase factors.
We assume that the initial function is sufficiently smooth such that a shallow depth PQC can accurately perform state preparation --- this was shown to be true in~\cite{lubasch2020, Plekhanov2022variationalquantum, marin2023quantum, akhalwaya2023modular} for sinusoidal, Gaussian and log-normal functions.

The diagonal evolution operator given in Eq.~\eqref{eq: diagonal operator} can be expressed as
\begin{align}\label{eq: diag op without Z}
    e^{-\text{i} t Z \otimes \sqrt{-D}}  &= 
    \sum_{k} \Big( e^{-\text{i} t 2 N \sin(\frac{k\pi}{N})} \ket{0}\bra{0} \nonumber \\
    &+ e^{\text{i} t 2 N \sin(\frac{k\pi}{N})}\ket{1}\bra{1} \Big) \otimes \ket{k}\bra{k}.
\end{align}
In the following we assume w.l.o.g. that the initial state is static such that the initial amplitudes in the velocity space are zero, i.e. $c_{1,k} = 0 \, \forall k$.
The exact time evolved state is therefore given by
\begin{align}\label{eq: diag on init}
\ket{\psi_{\text{exact}}} &= e^{-\text{i} t Z \otimes \sqrt{-D}} \ket{\psi_{\text{init}}}\nonumber\\
&= \sum_k e^{-\text{i} t 2 N \sin\left(\frac{k\pi}{N}\right)} c_{0,k} \ket{0,k}.
\end{align}
By making the approximation $\sin\left({\pi k}/{N}\right) \approx {\pi k}/{N}$ we compute the approximate time evolved state
\begin{align}\label{eq: approx state}
    \ket{\psi_{\text{approx}}} 
    &=\sum_k e^{-\text{i} t 2 k \pi} c_{0,k} \ket{0,k}. 
\end{align}
In order to compute the infidelity, we first calculate the overlap between the exact and approximate states
\begin{align}\label{eq: overlap exact approx}
 \bra{\psi_{\text{approx}}}\ket{\psi_{\text{exact}}}
 &= \sum_{k}|c_{0,k}|^2e^{-\text{i}t\left(2N\sin(\frac{k\pi}{N})-2k\pi\right)}.
\end{align}

We define $\alpha_{k} = 2 N \sin(k \pi / N) - 2 k \pi$ and assume that the initial state is sufficiently smooth such that $\alpha_{k}t\ll1$ for all times $t$ considered.
A Taylor approximation of the overlap Eq.~\eqref{eq: overlap exact approx} about $\alpha_{k}t = 0$ gives
\begin{align}\label{eq: approx overlap}\bra{\psi_{\text{approx}}}\ket{\psi_{\text{exact}}} &= \sum_{k}|c_{0,k}|^2\left[ 1 -\text{i}t\alpha_k - \frac{t^2\alpha_k^2}{2!} + O(t^3\alpha_k^3)\right].
\end{align}
The infidelity is approximated as
\begin{align}\label{eq: not expanded infidelity}
    \epsilon(t) &= 1 - |\bra{\psi_{\text{approx}}}\ket{\psi_{\text{exact}}}|^2 \nonumber \\
    &= \sum_{k}|c_{0,k}|^2t^2\alpha_k^2 - t^2\left(\sum_k|c_{0,k}|^2\alpha_{k}\right)^2
    +O(t^4\alpha_k^4).
\end{align}
Retaining the assumption of smooth initial conditions, we approximate $\alpha_{k}$ to low order in $|k|/N\ll1$, leading to an infidelity
\begin{align}
    \epsilon &\leq \frac{t^2\pi^6}{9N^4}\sum_{k}|c_{0,k}|^2\left(k^3-\sum_{k'}|c_{0,k'}|^{2}k'^3\right)^2.
\end{align}
Therefore the infidelity scales as $O(t^2 / N^4)$ which is consistent with what is observed in Fig.~\ref{fig: error vs N} and Fig.~\ref{fig: error vs t}.

\begin{figure}
     \centering
     \includegraphics[width=\linewidth]{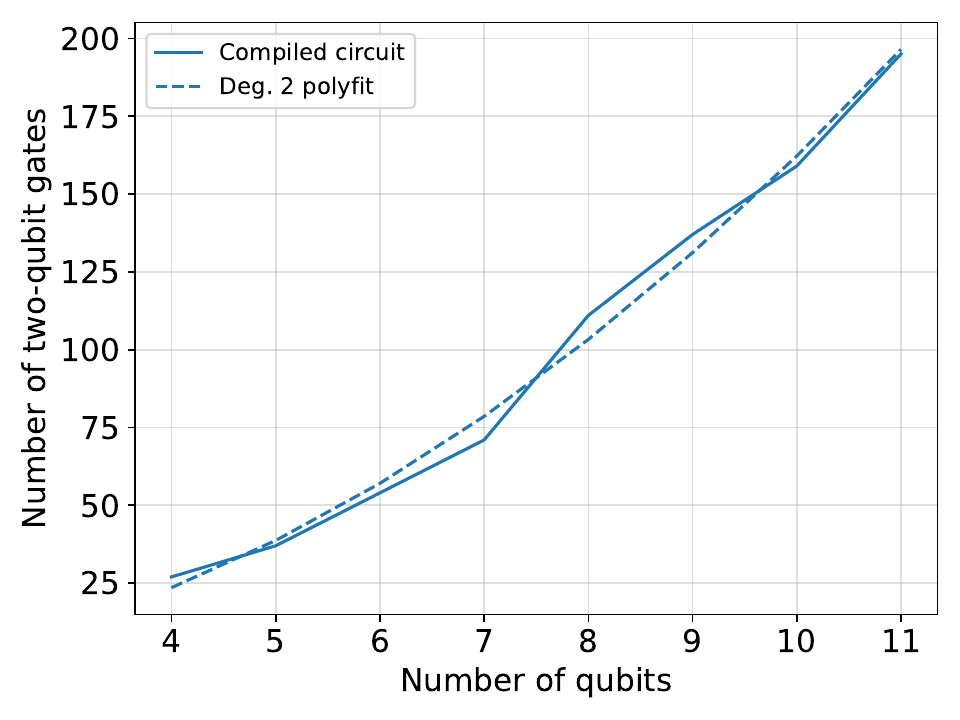}
     \caption{\label{fig: 2bit gatge vs n}
     Number of two-qubit gates from compiled circuits as a function of the number of qubits at $t = 1$ (solid line).
     Infidelity $\epsilon(t)$ is not fixed. Results are in line with a degree-2 polynomial fit (dashed) as expected from their asymptotic scaling of $O(n^2)$.
     The circuits were compiled in the series H1-1 gateset~\cite{H1datasheet} using Pytket with optimization level 2~\cite{sivarajah2020t}.
     }
\end{figure}

The number of two-qubit gates scales as $O(n^2)$ due to the exact application of the QFT.
This can be seen in Fig.~\ref{fig: 2bit gatge vs n} where we have utilized Pytket's quantum circuit compilation code \cite{sivarajah2020t}.
Additionally, since the number of grid points is $N = 2^n$, we find the gate complexity of our quantum algorithm is $O\left(\text{polylog}\left(\frac{t}{\sqrt{\epsilon}}\right)\right)$.
This sublinear scaling in $t$ classifies our approach as a `fast-forwarding' quantum algorithm, which is consistent with results of, e.g.~\cite{an2023theory}.

\bibliography{refs.bib} 

\begin{thebibliography}{70}%
\makeatletter
\providecommand \@ifxundefined [1]{%
 \@ifx{#1\undefined}
}%
\providecommand \@ifnum [1]{%
 \ifnum #1\expandafter \@firstoftwo
 \else \expandafter \@secondoftwo
 \fi
}%
\providecommand \@ifx [1]{%
 \ifx #1\expandafter \@firstoftwo
 \else \expandafter \@secondoftwo
 \fi
}%
\providecommand \natexlab [1]{#1}%
\providecommand \enquote  [1]{``#1''}%
\providecommand \bibnamefont  [1]{#1}%
\providecommand \bibfnamefont [1]{#1}%
\providecommand \citenamefont [1]{#1}%
\providecommand \href@noop [0]{\@secondoftwo}%
\providecommand \href [0]{\begingroup \@sanitize@url \@href}%
\providecommand \@href[1]{\@@startlink{#1}\@@href}%
\providecommand \@@href[1]{\endgroup#1\@@endlink}%
\providecommand \@sanitize@url [0]{\catcode `\\12\catcode `\$12\catcode
  `\&12\catcode `\#12\catcode `\^12\catcode `\_12\catcode `\%12\relax}%
\providecommand \@@startlink[1]{}%
\providecommand \@@endlink[0]{}%
\providecommand \url  [0]{\begingroup\@sanitize@url \@url }%
\providecommand \@url [1]{\endgroup\@href {#1}{\urlprefix }}%
\providecommand \urlprefix  [0]{URL }%
\providecommand \Eprint [0]{\href }%
\providecommand \doibase [0]{https://doi.org/}%
\providecommand \selectlanguage [0]{\@gobble}%
\providecommand \bibinfo  [0]{\@secondoftwo}%
\providecommand \bibfield  [0]{\@secondoftwo}%
\providecommand \translation [1]{[#1]}%
\providecommand \BibitemOpen [0]{}%
\providecommand \bibitemStop [0]{}%
\providecommand \bibitemNoStop [0]{.\EOS\space}%
\providecommand \EOS [0]{\spacefactor3000\relax}%
\providecommand \BibitemShut  [1]{\csname bibitem#1\endcsname}%
\let\auto@bib@innerbib\@empty
%</preamble>
\bibitem [{\citenamefont {Li}\ and\ \citenamefont {Qu}(2022)}]{li2022research}%
  \BibitemOpen
  \bibfield  {author} {\bibinfo {author} {\bibfnamefont {Z.-C.}\ \bibnamefont
  {Li}}\ and\ \bibinfo {author} {\bibfnamefont {Y.-M.}\ \bibnamefont {Qu}},\
  }\bibfield  {title} {\bibinfo {title} {Research progress on seismic imaging
  technology},\ }\href
  {https://doi.org/https://doi.org/10.1016/j.petsci.2022.01.015} {\bibfield
  {journal} {\bibinfo  {journal} {Pet. Sci.}\ }\textbf {\bibinfo {volume}
  {19}},\ \bibinfo {pages} {128} (\bibinfo {year} {2022})}\BibitemShut
  {NoStop}%
\bibitem [{\citenamefont {Zawawi}\ \emph {et~al.}(2018)\citenamefont {Zawawi},
  \citenamefont {Saleha}, \citenamefont {Salwa}, \citenamefont {Hassan},
  \citenamefont {Zahari}, \citenamefont {Ramli},\ and\ \citenamefont
  {Muda}}]{zawawi2018review}%
  \BibitemOpen
  \bibfield  {author} {\bibinfo {author} {\bibfnamefont {M.~H.}\ \bibnamefont
  {Zawawi}}, \bibinfo {author} {\bibfnamefont {A.}~\bibnamefont {Saleha}},
  \bibinfo {author} {\bibfnamefont {A.}~\bibnamefont {Salwa}}, \bibinfo
  {author} {\bibfnamefont {N.~H.}\ \bibnamefont {Hassan}}, \bibinfo {author}
  {\bibfnamefont {N.~M.}\ \bibnamefont {Zahari}}, \bibinfo {author}
  {\bibfnamefont {M.~Z.}\ \bibnamefont {Ramli}},\ and\ \bibinfo {author}
  {\bibfnamefont {Z.~C.}\ \bibnamefont {Muda}},\ }\bibfield  {title} {\bibinfo
  {title} {{A review: Fundamentals of computational fluid dynamics (CFD)}},\
  }\href {https://doi.org/10.1063/1.5066893} {\bibfield  {journal} {\bibinfo
  {journal} {AIP Conf. Proc.}\ }\textbf {\bibinfo {volume} {2030}},\ \bibinfo
  {pages} {020252} (\bibinfo {year} {2018})}\BibitemShut {NoStop}%
\bibitem [{\citenamefont {Campolieti}\ and\ \citenamefont
  {Makarov}(2022)}]{campolieti2018financial}%
  \BibitemOpen
  \bibfield  {author} {\bibinfo {author} {\bibfnamefont {G.}~\bibnamefont
  {Campolieti}}\ and\ \bibinfo {author} {\bibfnamefont {R.~N.}\ \bibnamefont
  {Makarov}},\ }\href {https://doi.org/10.1201/9780429468889} {\emph {\bibinfo
  {title} {{Financial Mathematics: A Comprehensive Treatment}}}}\ (\bibinfo
  {publisher} {in Continuous Time Volume II (1st ed.). Chapman and Hall/CRC},\
  \bibinfo {year} {2022})\BibitemShut {NoStop}%
\bibitem [{\citenamefont {Pople}(1999)}]{pople1999nobel}%
  \BibitemOpen
  \bibfield  {author} {\bibinfo {author} {\bibfnamefont {J.~A.}\ \bibnamefont
  {Pople}},\ }\bibfield  {title} {\bibinfo {title} {{Nobel Lecture: Quantum
  chemical models}},\ }\href {https://doi.org/10.1103/RevModPhys.71.1267}
  {\bibfield  {journal} {\bibinfo  {journal} {Rev. Mod. Phys.}\ }\textbf
  {\bibinfo {volume} {71}},\ \bibinfo {pages} {1267} (\bibinfo {year}
  {1999})}\BibitemShut {NoStop}%
\bibitem [{\citenamefont {McArdle}\ \emph {et~al.}(2020)\citenamefont
  {McArdle}, \citenamefont {Endo}, \citenamefont {Aspuru-Guzik}, \citenamefont
  {Benjamin},\ and\ \citenamefont {Yuan}}]{mcardle2020quantum}%
  \BibitemOpen
  \bibfield  {author} {\bibinfo {author} {\bibfnamefont {S.}~\bibnamefont
  {McArdle}}, \bibinfo {author} {\bibfnamefont {S.}~\bibnamefont {Endo}},
  \bibinfo {author} {\bibfnamefont {A.}~\bibnamefont {Aspuru-Guzik}}, \bibinfo
  {author} {\bibfnamefont {S.~C.}\ \bibnamefont {Benjamin}},\ and\ \bibinfo
  {author} {\bibfnamefont {X.}~\bibnamefont {Yuan}},\ }\bibfield  {title}
  {\bibinfo {title} {Quantum computational chemistry},\ }\href
  {https://doi.org/10.1103/RevModPhys.92.015003} {\bibfield  {journal}
  {\bibinfo  {journal} {Rev. Mod. Phys.}\ }\textbf {\bibinfo {volume} {92}},\
  \bibinfo {pages} {015003} (\bibinfo {year} {2020})}\BibitemShut {NoStop}%
\bibitem [{\citenamefont {Durran}(2013)}]{durran2013numerical}%
  \BibitemOpen
  \bibfield  {author} {\bibinfo {author} {\bibfnamefont {D.~R.}\ \bibnamefont
  {Durran}},\ }\href {https://doi.org/10.1007/978-1-4757-3081-4} {\emph
  {\bibinfo {title} {Numerical methods for wave equations in geophysical fluid
  dynamics}}},\ Vol.~\bibinfo {volume} {32}\ (\bibinfo  {publisher} {Springer
  Science \& Business Media},\ \bibinfo {year} {2013})\BibitemShut {NoStop}%
\bibitem [{\citenamefont {Alquran}\ \emph {et~al.}(2012)\citenamefont
  {Alquran}, \citenamefont {Ali},\ and\ \citenamefont
  {Al-Khaled}}]{alquran2012solitary}%
  \BibitemOpen
  \bibfield  {author} {\bibinfo {author} {\bibfnamefont {M.}~\bibnamefont
  {Alquran}}, \bibinfo {author} {\bibfnamefont {M.}~\bibnamefont {Ali}},\ and\
  \bibinfo {author} {\bibfnamefont {K.}~\bibnamefont {Al-Khaled}},\ }\bibfield
  {title} {\bibinfo {title} {Solitary wave solutions to shallow water waves
  arising in fluid dynamics},\ }\href
  {http://www.scopus.com/inward/record.url?scp=84872452939&partnerID=8YFLogxK}
  {\bibfield  {journal} {\bibinfo  {journal} {Nonlinear Stud.}\ }\textbf
  {\bibinfo {volume} {19}},\ \bibinfo {pages} {555} (\bibinfo {year}
  {2012})}\BibitemShut {NoStop}%
\bibitem [{\citenamefont
  {B{\'e}cherrawy}(2013)}]{becherrawy2013electromagnetism}%
  \BibitemOpen
  \bibfield  {author} {\bibinfo {author} {\bibfnamefont {T.}~\bibnamefont
  {B{\'e}cherrawy}},\ }\href
  {https://onlinelibrary.wiley.com/doi/abs/10.1002/9781118562215} {\emph
  {\bibinfo {title} {Electromagnetism: Maxwell equations, wave propagation and
  emission}}}\ (\bibinfo  {publisher} {John Wiley \& Sons},\ \bibinfo {year}
  {2013})\BibitemShut {NoStop}%
\bibitem [{\citenamefont {Chew}\ \emph {et~al.}(2022)\citenamefont {Chew},
  \citenamefont {Tong},\ and\ \citenamefont {Bin}}]{chew2022integral}%
  \BibitemOpen
  \bibfield  {author} {\bibinfo {author} {\bibfnamefont {W.}~\bibnamefont
  {Chew}}, \bibinfo {author} {\bibfnamefont {M.-S.}\ \bibnamefont {Tong}},\
  and\ \bibinfo {author} {\bibfnamefont {H.}~\bibnamefont {Bin}},\ }\href
  {https://doi.org/10.1007/978-3-031-01707-0} {\emph {\bibinfo {title}
  {Integral equation methods for electromagnetic and elastic waves}}}\
  (\bibinfo  {publisher} {Springer Nature},\ \bibinfo {year}
  {2022})\BibitemShut {NoStop}%
\bibitem [{\citenamefont {Broglie}(1924)}]{broglie1924xxxv}%
  \BibitemOpen
  \bibfield  {author} {\bibinfo {author} {\bibfnamefont {L.~d.}\ \bibnamefont
  {Broglie}},\ }\bibfield  {title} {\bibinfo {title} {A tentative theory of
  light quanta},\ }\href {https://doi.org/10.1080/14786442408634378} {\bibfield
   {journal} {\bibinfo  {journal} {London, Edinburgh Dublin Philos. Mag. J.
  Sci.}\ }\textbf {\bibinfo {volume} {47}},\ \bibinfo {pages} {446} (\bibinfo
  {year} {1924})}\BibitemShut {NoStop}%
\bibitem [{\citenamefont {Schr{\"o}dinger}(2003)}]{schrodinger2003collected}%
  \BibitemOpen
  \bibfield  {author} {\bibinfo {author} {\bibfnamefont {E.}~\bibnamefont
  {Schr{\"o}dinger}},\ }\href
  {https://books.google.co.uk/books?id=8ZROAgAAQBAJ} {\emph {\bibinfo {title}
  {Collected Papers on Wave Mechanics}}}\ (\bibinfo  {publisher} {AMS Chelsea
  Pub.},\ \bibinfo {year} {2003})\BibitemShut {NoStop}%
\bibitem [{\citenamefont {Virieux}\ and\ \citenamefont
  {Operto}(2009)}]{virieux2009fwi}%
  \BibitemOpen
  \bibfield  {author} {\bibinfo {author} {\bibfnamefont {J.}~\bibnamefont
  {Virieux}}\ and\ \bibinfo {author} {\bibfnamefont {S.}~\bibnamefont
  {Operto}},\ }\bibfield  {title} {\bibinfo {title} {An overview of
  full-waveform inversion in exploration geophysics},\ }\href
  {https://doi.org/10.1190/1.3238367} {\bibfield  {journal} {\bibinfo
  {journal} {Geophysics}\ }\textbf {\bibinfo {volume} {74}} (\bibinfo {year}
  {2009})}\BibitemShut {NoStop}%
\bibitem [{\citenamefont {Virieux}\ \emph {et~al.}(2017)\citenamefont
  {Virieux}, \citenamefont {Asnaashari}, \citenamefont {Brossier},
  \citenamefont {Métivier}, \citenamefont {Ribodetti},\ and\ \citenamefont
  {Zhou}}]{virieux2017introduction}%
  \BibitemOpen
  \bibfield  {author} {\bibinfo {author} {\bibfnamefont {J.}~\bibnamefont
  {Virieux}}, \bibinfo {author} {\bibfnamefont {A.}~\bibnamefont {Asnaashari}},
  \bibinfo {author} {\bibfnamefont {R.}~\bibnamefont {Brossier}}, \bibinfo
  {author} {\bibfnamefont {L.}~\bibnamefont {Métivier}}, \bibinfo {author}
  {\bibfnamefont {A.}~\bibnamefont {Ribodetti}},\ and\ \bibinfo {author}
  {\bibfnamefont {W.}~\bibnamefont {Zhou}},\ }\bibinfo {title} {An introduction
  to full waveform inversion},\ in\ \href
  {https://doi.org/10.1190/1.9781560803027.entry6} {\emph {\bibinfo {booktitle}
  {Encyclopedia of Exploration Geophysics}}}\ (\bibinfo {year} {2017})\ pp.\
  \bibinfo {pages} {R1--1--R1--40}\BibitemShut {NoStop}%
\bibitem [{\citenamefont {Brandsberg-Dahl}(2017)}]{brandsberg2017high}%
  \BibitemOpen
  \bibfield  {author} {\bibinfo {author} {\bibfnamefont {S.}~\bibnamefont
  {Brandsberg-Dahl}},\ }\bibinfo {title} {High-performance computing for
  seismic imaging; from shoestrings to the cloud},\ in\ \href
  {https://doi.org/10.1190/segam2017-17795566.1} {\emph {\bibinfo {booktitle}
  {SEG Tech. Program Expand. Abstr. 2017}}}\ (\bibinfo {year} {2017})\ pp.\
  \bibinfo {pages} {5273--5277}\BibitemShut {NoStop}%
\bibitem [{\citenamefont {Etgen}\ and\ \citenamefont
  {Dellinger}(2005)}]{etgen1989accurate}%
  \BibitemOpen
  \bibfield  {author} {\bibinfo {author} {\bibfnamefont {J.~T.}\ \bibnamefont
  {Etgen}}\ and\ \bibinfo {author} {\bibfnamefont {J.}~\bibnamefont
  {Dellinger}},\ }\bibinfo {title} {Accurate wave‐equation modeling},\ in\
  \href {https://doi.org/10.1190/1.1889673} {\emph {\bibinfo {booktitle} {SEG
  Tech. Program Expand. Abstr. 1989}}}\ (\bibinfo {year} {2005})\ pp.\ \bibinfo
  {pages} {494--497}\BibitemShut {NoStop}%
\bibitem [{\citenamefont {Etgen}\ and\ \citenamefont
  {O’Brien}(2007)}]{etgen2007computational}%
  \BibitemOpen
  \bibfield  {author} {\bibinfo {author} {\bibfnamefont {J.~T.}\ \bibnamefont
  {Etgen}}\ and\ \bibinfo {author} {\bibfnamefont {M.~J.}\ \bibnamefont
  {O’Brien}},\ }\bibfield  {title} {\bibinfo {title} {Computational methods
  for large-scale 3d acoustic finite-difference modeling: A tutorial},\ }\href
  {https://doi.org/10.1190/1.2753753} {\bibfield  {journal} {\bibinfo
  {journal} {Geophysics}\ }\textbf {\bibinfo {volume} {72}},\ \bibinfo {pages}
  {SM223} (\bibinfo {year} {2007})}\BibitemShut {NoStop}%
\bibitem [{\citenamefont {Fornberg}(1987)}]{fornberg1986seg}%
  \BibitemOpen
  \bibfield  {author} {\bibinfo {author} {\bibfnamefont {B.}~\bibnamefont
  {Fornberg}},\ }\bibfield  {title} {\bibinfo {title} {The pseudospectral
  method: Comparisons with finite differences for the elastic wave equation},\
  }\href {https://doi.org/10.1190/1.1892921} {\bibfield  {journal} {\bibinfo
  {journal} {Geophysics}\ }\textbf {\bibinfo {volume} {52}},\ \bibinfo {pages}
  {483–501} (\bibinfo {year} {1987})}\BibitemShut {NoStop}%
\bibitem [{\citenamefont {Zalka}(1998)}]{ZalkaSimulating1998}%
  \BibitemOpen
  \bibfield  {author} {\bibinfo {author} {\bibfnamefont {C.}~\bibnamefont
  {Zalka}},\ }\bibfield  {title} {\bibinfo {title} {Simulating quantum systems
  on a quantum computer},\ }\href {https://doi.org/10.1098/rspa.1998.0162}
  {\bibfield  {journal} {\bibinfo  {journal} {Proc. R. Soc. A: Math. Phys. Eng.
  Sci.}\ }\textbf {\bibinfo {volume} {454}},\ \bibinfo {pages} {313} (\bibinfo
  {year} {1998})}\BibitemShut {NoStop}%
\bibitem [{\citenamefont {Nielsen}\ and\ \citenamefont
  {Chuang}(2010)}]{ChuangQuantum2010}%
  \BibitemOpen
  \bibfield  {author} {\bibinfo {author} {\bibfnamefont {M.~A.}\ \bibnamefont
  {Nielsen}}\ and\ \bibinfo {author} {\bibfnamefont {I.~L.}\ \bibnamefont
  {Chuang}},\ }\href {https://doi.org/10.1017/CBO9780511976667} {\emph
  {\bibinfo {title} {{Quantum Computation and Quantum Information: 10th
  Anniversary Edition}}}}\ (\bibinfo  {publisher} {Cambridge University
  Press},\ \bibinfo {address} {Cambridge},\ \bibinfo {year} {2010})\BibitemShut
  {NoStop}%
\bibitem [{\citenamefont {Childs}\ \emph {et~al.}(2022)\citenamefont {Childs},
  \citenamefont {Leng}, \citenamefont {Li}, \citenamefont {Liu},\ and\
  \citenamefont {Zhang}}]{childs2022quantum}%
  \BibitemOpen
  \bibfield  {author} {\bibinfo {author} {\bibfnamefont {A.~M.}\ \bibnamefont
  {Childs}}, \bibinfo {author} {\bibfnamefont {J.}~\bibnamefont {Leng}},
  \bibinfo {author} {\bibfnamefont {T.}~\bibnamefont {Li}}, \bibinfo {author}
  {\bibfnamefont {J.-P.}\ \bibnamefont {Liu}},\ and\ \bibinfo {author}
  {\bibfnamefont {C.}~\bibnamefont {Zhang}},\ }\bibfield  {title} {\bibinfo
  {title} {Quantum simulation of real-space dynamics},\ }\href
  {https://doi.org/10.22331/q-2022-11-17-860} {\bibfield  {journal} {\bibinfo
  {journal} {{Quantum}}\ }\textbf {\bibinfo {volume} {6}},\ \bibinfo {pages}
  {860} (\bibinfo {year} {2022})}\BibitemShut {NoStop}%
\bibitem [{\citenamefont {Cao}\ \emph {et~al.}(2013)\citenamefont {Cao},
  \citenamefont {Papageorgiou}, \citenamefont {Petras}, \citenamefont {Traub},\
  and\ \citenamefont {Kais}}]{Cao2013Quantum}%
  \BibitemOpen
  \bibfield  {author} {\bibinfo {author} {\bibfnamefont {Y.}~\bibnamefont
  {Cao}}, \bibinfo {author} {\bibfnamefont {A.}~\bibnamefont {Papageorgiou}},
  \bibinfo {author} {\bibfnamefont {I.}~\bibnamefont {Petras}}, \bibinfo
  {author} {\bibfnamefont {J.}~\bibnamefont {Traub}},\ and\ \bibinfo {author}
  {\bibfnamefont {S.}~\bibnamefont {Kais}},\ }\bibfield  {title} {\bibinfo
  {title} {Quantum algorithm and circuit design solving the {Poisson}
  equation},\ }\href {https://doi.org/10.1088/1367-2630/15/1/013021} {\bibfield
   {journal} {\bibinfo  {journal} {New J. Phys.}\ }\textbf {\bibinfo {volume}
  {15}},\ \bibinfo {pages} {013021} (\bibinfo {year} {2013})}\BibitemShut
  {NoStop}%
\bibitem [{\citenamefont {Berry}(2014)}]{berry2014high}%
  \BibitemOpen
  \bibfield  {author} {\bibinfo {author} {\bibfnamefont {D.~W.}\ \bibnamefont
  {Berry}},\ }\bibfield  {title} {\bibinfo {title} {High-order quantum
  algorithm for solving linear differential equations},\ }\href
  {https://doi.org/10.1088/1751-8113/47/10/105301} {\bibfield  {journal}
  {\bibinfo  {journal} {J. Phys. A}\ }\textbf {\bibinfo {volume} {47}},\
  \bibinfo {pages} {105301} (\bibinfo {year} {2014})}\BibitemShut {NoStop}%
\bibitem [{\citenamefont {Berry}\ \emph {et~al.}(2017)\citenamefont {Berry},
  \citenamefont {Childs}, \citenamefont {Ostrander},\ and\ \citenamefont
  {Wang}}]{Berry2017Quantum}%
  \BibitemOpen
  \bibfield  {author} {\bibinfo {author} {\bibfnamefont {D.~W.}\ \bibnamefont
  {Berry}}, \bibinfo {author} {\bibfnamefont {A.~M.}\ \bibnamefont {Childs}},
  \bibinfo {author} {\bibfnamefont {A.}~\bibnamefont {Ostrander}},\ and\
  \bibinfo {author} {\bibfnamefont {G.}~\bibnamefont {Wang}},\ }\bibfield
  {title} {\bibinfo {title} {{Quantum Algorithm for Linear Differential
  Equations with Exponentially Improved Dependence on Precision}},\ }\href
  {https://doi.org/10.1007/s00220-017-3002-y} {\bibfield  {journal} {\bibinfo
  {journal} {Commun. Math. Phys.}\ }\textbf {\bibinfo {volume} {356}},\
  \bibinfo {pages} {1057} (\bibinfo {year} {2017})}\BibitemShut {NoStop}%
\bibitem [{\citenamefont {Childs}\ \emph {et~al.}(2017)\citenamefont {Childs},
  \citenamefont {Kothari},\ and\ \citenamefont {Somma}}]{childs2017quantum}%
  \BibitemOpen
  \bibfield  {author} {\bibinfo {author} {\bibfnamefont {A.~M.}\ \bibnamefont
  {Childs}}, \bibinfo {author} {\bibfnamefont {R.}~\bibnamefont {Kothari}},\
  and\ \bibinfo {author} {\bibfnamefont {R.~D.}\ \bibnamefont {Somma}},\
  }\bibfield  {title} {\bibinfo {title} {Quantum algorithm for systems of
  linear equations with exponentially improved dependence on precision},\
  }\href {https://doi.org/10.1137/16M1087072} {\bibfield  {journal} {\bibinfo
  {journal} {SIAM J. Comput.}\ }\textbf {\bibinfo {volume} {46}},\ \bibinfo
  {pages} {1920} (\bibinfo {year} {2017})}\BibitemShut {NoStop}%
\bibitem [{\citenamefont {Lloyd}\ \emph {et~al.}(2020)\citenamefont {Lloyd},
  \citenamefont {De~Palma}, \citenamefont {Gokler}, \citenamefont {Kiani},
  \citenamefont {Liu}, \citenamefont {Marvian}, \citenamefont {Tennie},\ and\
  \citenamefont {Palmer}}]{lloyd2020quantum}%
  \BibitemOpen
  \bibfield  {author} {\bibinfo {author} {\bibfnamefont {S.}~\bibnamefont
  {Lloyd}}, \bibinfo {author} {\bibfnamefont {G.}~\bibnamefont {De~Palma}},
  \bibinfo {author} {\bibfnamefont {C.}~\bibnamefont {Gokler}}, \bibinfo
  {author} {\bibfnamefont {B.}~\bibnamefont {Kiani}}, \bibinfo {author}
  {\bibfnamefont {Z.-W.}\ \bibnamefont {Liu}}, \bibinfo {author} {\bibfnamefont
  {M.}~\bibnamefont {Marvian}}, \bibinfo {author} {\bibfnamefont
  {F.}~\bibnamefont {Tennie}},\ and\ \bibinfo {author} {\bibfnamefont
  {T.}~\bibnamefont {Palmer}},\ }\bibfield  {title} {\bibinfo {title} {Quantum
  algorithm for nonlinear differential equations},\ }\Eprint
  {https://arxiv.org/abs/2011.06571} {arXiv:2011.06571 [quant-ph]}  (\bibinfo
  {year} {2020})\BibitemShut {NoStop}%
\bibitem [{\citenamefont {Childs}\ \emph {et~al.}(2021)\citenamefont {Childs},
  \citenamefont {Liu},\ and\ \citenamefont {Ostrander}}]{childs2021high}%
  \BibitemOpen
  \bibfield  {author} {\bibinfo {author} {\bibfnamefont {A.~M.}\ \bibnamefont
  {Childs}}, \bibinfo {author} {\bibfnamefont {J.-P.}\ \bibnamefont {Liu}},\
  and\ \bibinfo {author} {\bibfnamefont {A.}~\bibnamefont {Ostrander}},\
  }\bibfield  {title} {\bibinfo {title} {High-precision quantum algorithms for
  partial differential equations},\ }\href
  {https://doi.org/10.22331/q-2021-11-10-574} {\bibfield  {journal} {\bibinfo
  {journal} {Quantum}\ }\textbf {\bibinfo {volume} {5}},\ \bibinfo {pages}
  {574} (\bibinfo {year} {2021})}\BibitemShut {NoStop}%
\bibitem [{\citenamefont {Krovi}(2023)}]{Krovi2023improvedquantum}%
  \BibitemOpen
  \bibfield  {author} {\bibinfo {author} {\bibfnamefont {H.}~\bibnamefont
  {Krovi}},\ }\bibfield  {title} {\bibinfo {title} {Improved quantum algorithms
  for linear and nonlinear differential equations},\ }\href
  {https://doi.org/10.22331/q-2023-02-02-913} {\bibfield  {journal} {\bibinfo
  {journal} {{Quantum}}\ }\textbf {\bibinfo {volume} {7}},\ \bibinfo {pages}
  {913} (\bibinfo {year} {2023})}\BibitemShut {NoStop}%
\bibitem [{\citenamefont {Liu}\ \emph {et~al.}(2021{\natexlab{a}})\citenamefont
  {Liu}, \citenamefont {Kolden}, \citenamefont {Krovi}, \citenamefont
  {Loureiro}, \citenamefont {Trivisa},\ and\ \citenamefont
  {Childs}}]{liu2021efficient}%
  \BibitemOpen
  \bibfield  {author} {\bibinfo {author} {\bibfnamefont {J.-P.}\ \bibnamefont
  {Liu}}, \bibinfo {author} {\bibfnamefont {H.~{\O}.}\ \bibnamefont {Kolden}},
  \bibinfo {author} {\bibfnamefont {H.~K.}\ \bibnamefont {Krovi}}, \bibinfo
  {author} {\bibfnamefont {N.~F.}\ \bibnamefont {Loureiro}}, \bibinfo {author}
  {\bibfnamefont {K.}~\bibnamefont {Trivisa}},\ and\ \bibinfo {author}
  {\bibfnamefont {A.~M.}\ \bibnamefont {Childs}},\ }\bibfield  {title}
  {\bibinfo {title} {Efficient quantum algorithm for dissipative nonlinear
  differential equations},\ }\href {https://doi.org/10.1073/pnas.2026805118}
  {\bibfield  {journal} {\bibinfo  {journal} {PNAS}\ }\textbf {\bibinfo
  {volume} {118}},\ \bibinfo {pages} {e2026805118} (\bibinfo {year}
  {2021}{\natexlab{a}})}\BibitemShut {NoStop}%
\bibitem [{\citenamefont {Wiesner}(1996)}]{Wiesner1996Simulations}%
  \BibitemOpen
  \bibfield  {author} {\bibinfo {author} {\bibfnamefont {S.}~\bibnamefont
  {Wiesner}},\ }\bibfield  {title} {\bibinfo {title} {Simulations of many-body
  quantum systems by a quantum computer},\ }\Eprint
  {https://arxiv.org/abs/quant-ph/9603028} {arXiv:quant-ph/9603028}  (\bibinfo
  {year} {1996})\BibitemShut {NoStop}%
\bibitem [{\citenamefont {Costa}\ \emph {et~al.}(2019)\citenamefont {Costa},
  \citenamefont {Jordan},\ and\ \citenamefont {Ostrander}}]{costa2019quantum}%
  \BibitemOpen
  \bibfield  {author} {\bibinfo {author} {\bibfnamefont {P.~C.~S.}\
  \bibnamefont {Costa}}, \bibinfo {author} {\bibfnamefont {S.}~\bibnamefont
  {Jordan}},\ and\ \bibinfo {author} {\bibfnamefont {A.}~\bibnamefont
  {Ostrander}},\ }\bibfield  {title} {\bibinfo {title} {Quantum algorithm for
  simulating the wave equation},\ }\href
  {https://doi.org/10.1103/PhysRevA.99.012323} {\bibfield  {journal} {\bibinfo
  {journal} {Phys. Rev. A}\ }\textbf {\bibinfo {volume} {99}},\ \bibinfo
  {pages} {012323} (\bibinfo {year} {2019})}\BibitemShut {NoStop}%
\bibitem [{\citenamefont {Arrazola}\ \emph {et~al.}(2019)\citenamefont
  {Arrazola}, \citenamefont {Kalajdzievski}, \citenamefont {Weedbrook},\ and\
  \citenamefont {Lloyd}}]{arrazola2019quantum}%
  \BibitemOpen
  \bibfield  {author} {\bibinfo {author} {\bibfnamefont {J.~M.}\ \bibnamefont
  {Arrazola}}, \bibinfo {author} {\bibfnamefont {T.}~\bibnamefont
  {Kalajdzievski}}, \bibinfo {author} {\bibfnamefont {C.}~\bibnamefont
  {Weedbrook}},\ and\ \bibinfo {author} {\bibfnamefont {S.}~\bibnamefont
  {Lloyd}},\ }\bibfield  {title} {\bibinfo {title} {Quantum algorithm for
  nonhomogeneous linear partial differential equations},\ }\href
  {https://doi.org/10.1103/PhysRevA.100.032306} {\bibfield  {journal} {\bibinfo
   {journal} {Phys. Rev. A}\ }\textbf {\bibinfo {volume} {100}},\ \bibinfo
  {pages} {032306} (\bibinfo {year} {2019})}\BibitemShut {NoStop}%
\bibitem [{\citenamefont {Jin}\ \emph {et~al.}(2022)\citenamefont {Jin},
  \citenamefont {Liu},\ and\ \citenamefont {Yu}}]{jin2022quantum}%
  \BibitemOpen
  \bibfield  {author} {\bibinfo {author} {\bibfnamefont {S.}~\bibnamefont
  {Jin}}, \bibinfo {author} {\bibfnamefont {N.}~\bibnamefont {Liu}},\ and\
  \bibinfo {author} {\bibfnamefont {Y.}~\bibnamefont {Yu}},\ }\href@noop {}
  {\bibinfo {title} {Quantum simulation of partial differential equations via
  {Schrodingerisation}}} (\bibinfo {year} {2022}),\ \Eprint
  {https://arxiv.org/abs/2212.13969} {arXiv:2212.13969 [quant-ph]} \BibitemShut
  {NoStop}%
\bibitem [{\citenamefont {Babbush}\ \emph {et~al.}(2023)\citenamefont
  {Babbush}, \citenamefont {Berry}, \citenamefont {Kothari}, \citenamefont
  {Somma},\ and\ \citenamefont {Wiebe}}]{babbush2023}%
  \BibitemOpen
  \bibfield  {author} {\bibinfo {author} {\bibfnamefont {R.}~\bibnamefont
  {Babbush}}, \bibinfo {author} {\bibfnamefont {D.~W.}\ \bibnamefont {Berry}},
  \bibinfo {author} {\bibfnamefont {R.}~\bibnamefont {Kothari}}, \bibinfo
  {author} {\bibfnamefont {R.~D.}\ \bibnamefont {Somma}},\ and\ \bibinfo
  {author} {\bibfnamefont {N.}~\bibnamefont {Wiebe}},\ }\bibfield  {title}
  {\bibinfo {title} {{Exponential Quantum Speedup in Simulating Coupled
  Classical Oscillators}},\ }\href {https://doi.org/10.1103/PhysRevX.13.041041}
  {\bibfield  {journal} {\bibinfo  {journal} {Phys. Rev. X}\ }\textbf {\bibinfo
  {volume} {13}},\ \bibinfo {pages} {041041} (\bibinfo {year}
  {2023})}\BibitemShut {NoStop}%
\bibitem [{\citenamefont {Xu}\ \emph {et~al.}(2018)\citenamefont {Xu},
  \citenamefont {Daley}, \citenamefont {Givi},\ and\ \citenamefont
  {Somma}}]{XuEtAl18}%
  \BibitemOpen
  \bibfield  {author} {\bibinfo {author} {\bibfnamefont {G.}~\bibnamefont
  {Xu}}, \bibinfo {author} {\bibfnamefont {A.~J.}\ \bibnamefont {Daley}},
  \bibinfo {author} {\bibfnamefont {P.}~\bibnamefont {Givi}},\ and\ \bibinfo
  {author} {\bibfnamefont {R.~D.}\ \bibnamefont {Somma}},\ }\bibfield  {title}
  {\bibinfo {title} {{Turbulent Mixing Simulation via a Quantum Algorithm}},\
  }\href {https://doi.org/10.2514/1.J055896} {\bibfield  {journal} {\bibinfo
  {journal} {AIAA J.}\ }\textbf {\bibinfo {volume} {56}},\ \bibinfo {pages}
  {687} (\bibinfo {year} {2018})}\BibitemShut {NoStop}%
\bibitem [{\citenamefont {Xu}\ \emph {et~al.}(2019)\citenamefont {Xu},
  \citenamefont {Daley}, \citenamefont {Givi},\ and\ \citenamefont
  {Somma}}]{XuEtAl19}%
  \BibitemOpen
  \bibfield  {author} {\bibinfo {author} {\bibfnamefont {G.}~\bibnamefont
  {Xu}}, \bibinfo {author} {\bibfnamefont {A.~J.}\ \bibnamefont {Daley}},
  \bibinfo {author} {\bibfnamefont {P.}~\bibnamefont {Givi}},\ and\ \bibinfo
  {author} {\bibfnamefont {R.~D.}\ \bibnamefont {Somma}},\ }\bibfield  {title}
  {\bibinfo {title} {Quantum algorithm for the computation of the reactant
  conversion rate in homogeneous turbulence},\ }\href
  {https://doi.org/10.1080/13647830.2019.1626025} {\bibfield  {journal}
  {\bibinfo  {journal} {Combust. Theory Model.}\ }\textbf {\bibinfo {volume}
  {23}},\ \bibinfo {pages} {1090} (\bibinfo {year} {2019})}\BibitemShut
  {NoStop}%
\bibitem [{\citenamefont {Givi}\ \emph {et~al.}(2020)\citenamefont {Givi},
  \citenamefont {Daley}, \citenamefont {Mavriplis},\ and\ \citenamefont
  {Malik}}]{GiviEtAl20}%
  \BibitemOpen
  \bibfield  {author} {\bibinfo {author} {\bibfnamefont {P.}~\bibnamefont
  {Givi}}, \bibinfo {author} {\bibfnamefont {A.~J.}\ \bibnamefont {Daley}},
  \bibinfo {author} {\bibfnamefont {D.}~\bibnamefont {Mavriplis}},\ and\
  \bibinfo {author} {\bibfnamefont {M.}~\bibnamefont {Malik}},\ }\bibfield
  {title} {\bibinfo {title} {{Quantum Speedup for Aeroscience and
  Engineering}},\ }\href {https://doi.org/10.2514/1.J059183} {\bibfield
  {journal} {\bibinfo  {journal} {AIAA Journal}\ }\textbf {\bibinfo {volume}
  {58}},\ \bibinfo {pages} {3715} (\bibinfo {year} {2020})}\BibitemShut
  {NoStop}%
\bibitem [{\citenamefont {Lubasch}\ \emph {et~al.}(2020)\citenamefont
  {Lubasch}, \citenamefont {Joo}, \citenamefont {Moinier}, \citenamefont
  {Kiffner},\ and\ \citenamefont {Jaksch}}]{lubasch2020}%
  \BibitemOpen
  \bibfield  {author} {\bibinfo {author} {\bibfnamefont {M.}~\bibnamefont
  {Lubasch}}, \bibinfo {author} {\bibfnamefont {J.}~\bibnamefont {Joo}},
  \bibinfo {author} {\bibfnamefont {P.}~\bibnamefont {Moinier}}, \bibinfo
  {author} {\bibfnamefont {M.}~\bibnamefont {Kiffner}},\ and\ \bibinfo {author}
  {\bibfnamefont {D.}~\bibnamefont {Jaksch}},\ }\bibfield  {title} {\bibinfo
  {title} {Variational quantum algorithms for nonlinear problems},\ }\href
  {https://doi.org/10.1103/PhysRevA.101.010301} {\bibfield  {journal} {\bibinfo
   {journal} {Phys. Rev. A}\ }\textbf {\bibinfo {volume} {101}},\ \bibinfo
  {pages} {010301} (\bibinfo {year} {2020})}\BibitemShut {NoStop}%
\bibitem [{\citenamefont {Liu}\ \emph {et~al.}(2021{\natexlab{b}})\citenamefont
  {Liu}, \citenamefont {Wu}, \citenamefont {Wan}, \citenamefont {Pan},
  \citenamefont {Qin}, \citenamefont {Gao},\ and\ \citenamefont
  {Wen}}]{liu2021}%
  \BibitemOpen
  \bibfield  {author} {\bibinfo {author} {\bibfnamefont {H.-L.}\ \bibnamefont
  {Liu}}, \bibinfo {author} {\bibfnamefont {Y.-S.}\ \bibnamefont {Wu}},
  \bibinfo {author} {\bibfnamefont {L.-C.}\ \bibnamefont {Wan}}, \bibinfo
  {author} {\bibfnamefont {S.-J.}\ \bibnamefont {Pan}}, \bibinfo {author}
  {\bibfnamefont {S.-J.}\ \bibnamefont {Qin}}, \bibinfo {author} {\bibfnamefont
  {F.}~\bibnamefont {Gao}},\ and\ \bibinfo {author} {\bibfnamefont {Q.-Y.}\
  \bibnamefont {Wen}},\ }\bibfield  {title} {\bibinfo {title} {Variational
  quantum algorithm for the {Poisson} equation},\ }\href
  {https://doi.org/10.1103/PhysRevA.104.022418} {\bibfield  {journal} {\bibinfo
   {journal} {Phys. Rev. A}\ }\textbf {\bibinfo {volume} {104}},\ \bibinfo
  {pages} {022418} (\bibinfo {year} {2021}{\natexlab{b}})}\BibitemShut
  {NoStop}%
\bibitem [{\citenamefont {Sato}\ \emph {et~al.}(2021)\citenamefont {Sato},
  \citenamefont {Kondo}, \citenamefont {Koide}, \citenamefont {Takamatsu},\
  and\ \citenamefont {Imoto}}]{sato2021}%
  \BibitemOpen
  \bibfield  {author} {\bibinfo {author} {\bibfnamefont {Y.}~\bibnamefont
  {Sato}}, \bibinfo {author} {\bibfnamefont {R.}~\bibnamefont {Kondo}},
  \bibinfo {author} {\bibfnamefont {S.}~\bibnamefont {Koide}}, \bibinfo
  {author} {\bibfnamefont {H.}~\bibnamefont {Takamatsu}},\ and\ \bibinfo
  {author} {\bibfnamefont {N.}~\bibnamefont {Imoto}},\ }\bibfield  {title}
  {\bibinfo {title} {Variational quantum algorithm based on the minimum
  potential energy for solving the {Poisson} equation},\ }\href
  {https://doi.org/10.1103/PhysRevA.104.052409} {\bibfield  {journal} {\bibinfo
   {journal} {Phys. Rev. A}\ }\textbf {\bibinfo {volume} {104}},\ \bibinfo
  {pages} {052409} (\bibinfo {year} {2021})}\BibitemShut {NoStop}%
\bibitem [{\citenamefont {Kyriienko}\ \emph {et~al.}(2021)\citenamefont
  {Kyriienko}, \citenamefont {Paine},\ and\ \citenamefont
  {Elfving}}]{kyriienko2021solving}%
  \BibitemOpen
  \bibfield  {author} {\bibinfo {author} {\bibfnamefont {O.}~\bibnamefont
  {Kyriienko}}, \bibinfo {author} {\bibfnamefont {A.~E.}\ \bibnamefont
  {Paine}},\ and\ \bibinfo {author} {\bibfnamefont {V.~E.}\ \bibnamefont
  {Elfving}},\ }\bibfield  {title} {\bibinfo {title} {Solving nonlinear
  differential equations with differentiable quantum circuits},\ }\href
  {https://doi.org/10.1103/PhysRevA.103.052416} {\bibfield  {journal} {\bibinfo
   {journal} {Phys. Rev. A}\ }\textbf {\bibinfo {volume} {103}},\ \bibinfo
  {pages} {052416} (\bibinfo {year} {2021})}\BibitemShut {NoStop}%
\bibitem [{\citenamefont {Joo}\ and\ \citenamefont
  {Moon}(2021)}]{joo2021quantum}%
  \BibitemOpen
  \bibfield  {author} {\bibinfo {author} {\bibfnamefont {J.}~\bibnamefont
  {Joo}}\ and\ \bibinfo {author} {\bibfnamefont {H.}~\bibnamefont {Moon}},\
  }\bibfield  {title} {\bibinfo {title} {Quantum variational {PDE} solver with
  machine learning},\ }\Eprint {https://arxiv.org/abs/2109.09216}
  {arXiv:2109.09216 [quant-ph]}  (\bibinfo {year} {2021})\BibitemShut {NoStop}%
\bibitem [{\citenamefont {Budinski}(2021)}]{budinski2021quantum}%
  \BibitemOpen
  \bibfield  {author} {\bibinfo {author} {\bibfnamefont {L.}~\bibnamefont
  {Budinski}},\ }\bibfield  {title} {\bibinfo {title} {Quantum algorithm for
  the advection--diffusion equation simulated with the lattice {Boltzmann}
  method},\ }\href {https://doi.org/10.1007/s11128-021-02996-3} {\bibfield
  {journal} {\bibinfo  {journal} {Quantum Inf. Process.}\ }\textbf {\bibinfo
  {volume} {20}},\ \bibinfo {pages} {57} (\bibinfo {year} {2021})}\BibitemShut
  {NoStop}%
\bibitem [{\citenamefont {Albino}\ \emph {et~al.}(2022)\citenamefont {Albino},
  \citenamefont {Jardim}, \citenamefont {Knupp}, \citenamefont {Neto},
  \citenamefont {Pires},\ and\ \citenamefont {Nascimento}}]{albino2022solving}%
  \BibitemOpen
  \bibfield  {author} {\bibinfo {author} {\bibfnamefont {A.~S.}\ \bibnamefont
  {Albino}}, \bibinfo {author} {\bibfnamefont {L.~C.}\ \bibnamefont {Jardim}},
  \bibinfo {author} {\bibfnamefont {D.~C.}\ \bibnamefont {Knupp}}, \bibinfo
  {author} {\bibfnamefont {A.~J.~S.}\ \bibnamefont {Neto}}, \bibinfo {author}
  {\bibfnamefont {O.~M.}\ \bibnamefont {Pires}},\ and\ \bibinfo {author}
  {\bibfnamefont {E.~G.~S.}\ \bibnamefont {Nascimento}},\ }\bibfield  {title}
  {\bibinfo {title} {Solving partial differential equations on near-term
  quantum computers},\ }\Eprint {https://arxiv.org/abs/2208.05805}
  {arXiv:2208.05805 [quant-ph]}  (\bibinfo {year} {2022})\BibitemShut {NoStop}%
\bibitem [{\citenamefont {Guseynov}\ \emph {et~al.}(2023)\citenamefont
  {Guseynov}, \citenamefont {Zhukov}, \citenamefont {Pogosov},\ and\
  \citenamefont {Lebedev}}]{Guseynov2023Depth}%
  \BibitemOpen
  \bibfield  {author} {\bibinfo {author} {\bibfnamefont {N.~M.}\ \bibnamefont
  {Guseynov}}, \bibinfo {author} {\bibfnamefont {A.~A.}\ \bibnamefont
  {Zhukov}}, \bibinfo {author} {\bibfnamefont {W.~V.}\ \bibnamefont
  {Pogosov}},\ and\ \bibinfo {author} {\bibfnamefont {A.~V.}\ \bibnamefont
  {Lebedev}},\ }\bibfield  {title} {\bibinfo {title} {Depth analysis of
  variational quantum algorithms for the heat equation},\ }\href
  {https://doi.org/10.1103/PhysRevA.107.052422} {\bibfield  {journal} {\bibinfo
   {journal} {Phys. Rev. A}\ }\textbf {\bibinfo {volume} {107}},\ \bibinfo
  {pages} {052422} (\bibinfo {year} {2023})}\BibitemShut {NoStop}%
\bibitem [{\citenamefont {Liu}\ \emph {et~al.}(2023)\citenamefont {Liu},
  \citenamefont {Chen}, \citenamefont {Shu}, \citenamefont {Rebentrost},
  \citenamefont {Liu}, \citenamefont {Chew}, \citenamefont {Khoo},\ and\
  \citenamefont {Cui}}]{liu2023variational}%
  \BibitemOpen
  \bibfield  {author} {\bibinfo {author} {\bibfnamefont {Y.}~\bibnamefont
  {Liu}}, \bibinfo {author} {\bibfnamefont {Z.}~\bibnamefont {Chen}}, \bibinfo
  {author} {\bibfnamefont {C.}~\bibnamefont {Shu}}, \bibinfo {author}
  {\bibfnamefont {P.}~\bibnamefont {Rebentrost}}, \bibinfo {author}
  {\bibfnamefont {Y.}~\bibnamefont {Liu}}, \bibinfo {author} {\bibfnamefont
  {S.~C.}\ \bibnamefont {Chew}}, \bibinfo {author} {\bibfnamefont {B.~C.}\
  \bibnamefont {Khoo}},\ and\ \bibinfo {author} {\bibfnamefont {Y.~D.}\
  \bibnamefont {Cui}},\ }\bibfield  {title} {\bibinfo {title} {A variational
  quantum algorithm-based numerical method for solving potential and {Stokes}
  flows},\ }\Eprint {https://arxiv.org/abs/2303.01805} {arXiv:2303.01805
  [physics.flu-dyn]}  (\bibinfo {year} {2023})\BibitemShut {NoStop}%
\bibitem [{\citenamefont {Jaksch}\ \emph {et~al.}(2023)\citenamefont {Jaksch},
  \citenamefont {Givi}, \citenamefont {Daley},\ and\ \citenamefont
  {Rung}}]{JakschEtAl23}%
  \BibitemOpen
  \bibfield  {author} {\bibinfo {author} {\bibfnamefont {D.}~\bibnamefont
  {Jaksch}}, \bibinfo {author} {\bibfnamefont {P.}~\bibnamefont {Givi}},
  \bibinfo {author} {\bibfnamefont {A.~J.}\ \bibnamefont {Daley}},\ and\
  \bibinfo {author} {\bibfnamefont {T.}~\bibnamefont {Rung}},\ }\bibfield
  {title} {\bibinfo {title} {{Variational Quantum Algorithms for Computational
  Fluid Dynamics}},\ }\href {https://doi.org/10.2514/1.J062426} {\bibfield
  {journal} {\bibinfo  {journal} {AIAA J.}\ }\textbf {\bibinfo {volume} {61}},\
  \bibinfo {pages} {1885} (\bibinfo {year} {2023})}\BibitemShut {NoStop}%
\bibitem [{\citenamefont {Preskill}(2018)}]{preskill2018quantum}%
  \BibitemOpen
  \bibfield  {author} {\bibinfo {author} {\bibfnamefont {J.}~\bibnamefont
  {Preskill}},\ }\bibfield  {title} {\bibinfo {title} {Quantum computing in the
  {NISQ} era and beyond},\ }\href {https://doi.org/10.22331/q-2018-08-06-79}
  {\bibfield  {journal} {\bibinfo  {journal} {Quantum}\ }\textbf {\bibinfo
  {volume} {2}},\ \bibinfo {pages} {79} (\bibinfo {year} {2018})}\BibitemShut
  {NoStop}%
\bibitem [{\citenamefont {Berry}\ \emph {et~al.}(2015)\citenamefont {Berry},
  \citenamefont {Childs},\ and\ \citenamefont {Kothari}}]{Berry2015a}%
  \BibitemOpen
  \bibfield  {author} {\bibinfo {author} {\bibfnamefont {D.~W.}\ \bibnamefont
  {Berry}}, \bibinfo {author} {\bibfnamefont {A.~M.}\ \bibnamefont {Childs}},\
  and\ \bibinfo {author} {\bibfnamefont {R.}~\bibnamefont {Kothari}},\
  }\bibfield  {title} {\bibinfo {title} {Hamiltonian simulation with nearly
  optimal dependence on all parameters},\ }in\ \href
  {https://doi.org/10.1109/FOCS.2015.54} {\emph {\bibinfo {booktitle} {2015
  IEEE 56th Annual Symposium on Foundations of Computer Science}}}\ (\bibinfo
  {year} {2015})\ pp.\ \bibinfo {pages} {792--809}\BibitemShut {NoStop}%
\bibitem [{\citenamefont {Suau}\ \emph {et~al.}(2021)\citenamefont {Suau},
  \citenamefont {Staffelbach},\ and\ \citenamefont
  {Calandra}}]{suau2021practical}%
  \BibitemOpen
  \bibfield  {author} {\bibinfo {author} {\bibfnamefont {A.}~\bibnamefont
  {Suau}}, \bibinfo {author} {\bibfnamefont {G.}~\bibnamefont {Staffelbach}},\
  and\ \bibinfo {author} {\bibfnamefont {H.}~\bibnamefont {Calandra}},\
  }\bibfield  {title} {\bibinfo {title} {{Practical Quantum Computing: Solving
  the Wave Equation Using a Quantum Approach}},\ }\bibfield  {journal}
  {\bibinfo  {journal} {ACM Transactions on Quantum Computing}\ }\textbf
  {\bibinfo {volume} {2}},\ \href {https://doi.org/10.1145/3430030}
  {10.1145/3430030} (\bibinfo {year} {2021})\BibitemShut {NoStop}%
\bibitem [{\citenamefont {Ricker}(1953)}]{ricker1953form}%
  \BibitemOpen
  \bibfield  {author} {\bibinfo {author} {\bibfnamefont {N.}~\bibnamefont
  {Ricker}},\ }\bibfield  {title} {\bibinfo {title} {The form and laws of
  propagation of seismic wavelets},\ }\href {https://doi.org/10.1190/1.1437843}
  {\bibfield  {journal} {\bibinfo  {journal} {Geophysics}\ }\textbf {\bibinfo
  {volume} {18}},\ \bibinfo {pages} {10} (\bibinfo {year} {1953})}\BibitemShut
  {NoStop}%
\bibitem [{\citenamefont {Wang}(2015)}]{wang2015generalized}%
  \BibitemOpen
  \bibfield  {author} {\bibinfo {author} {\bibfnamefont {Y.}~\bibnamefont
  {Wang}},\ }\bibfield  {title} {\bibinfo {title} {Generalized seismic
  wavelets},\ }\href {https://doi.org/10.1093/gji/ggv346} {\bibfield  {journal}
  {\bibinfo  {journal} {Geophys. J. Int.}\ }\textbf {\bibinfo {volume} {203}},\
  \bibinfo {pages} {1172} (\bibinfo {year} {2015})}\BibitemShut {NoStop}%
\bibitem [{H1d()}]{H1datasheet}%
  \BibitemOpen
  \href {https://www.quantinuum.com/hardware/h1} {\bibinfo {title} {{System
  Model H1 Powered by Honeywell}}},\ \bibinfo {note} {accessed on December
  1st-6th, 2023}\BibitemShut {NoStop}%
\bibitem [{\citenamefont {Press}\ \emph {et~al.}(1992)\citenamefont {Press},
  \citenamefont {Teukolsky}, \citenamefont {Vetterling},\ and\ \citenamefont
  {Flannery}}]{NumericalRecipes}%
  \BibitemOpen
  \bibfield  {author} {\bibinfo {author} {\bibfnamefont {W.~H.}\ \bibnamefont
  {Press}}, \bibinfo {author} {\bibfnamefont {S.~A.}\ \bibnamefont
  {Teukolsky}}, \bibinfo {author} {\bibfnamefont {W.~T.}\ \bibnamefont
  {Vetterling}},\ and\ \bibinfo {author} {\bibfnamefont {B.~P.}\ \bibnamefont
  {Flannery}},\ }\href
  {https://www.cambridge.org/us/universitypress/subjects/mathematics/numerical-recipes/numerical-recipes-example-book-c-2nd-edition?format=PB&isbn=9780521437202}
  {\emph {\bibinfo {title} {{Numerical Recipes in C}}}}\ (\bibinfo  {publisher}
  {Cambridge University Press},\ \bibinfo {address} {New York},\ \bibinfo
  {year} {1992})\BibitemShut {NoStop}%
\bibitem [{\citenamefont {Kikuchi}\ \emph {et~al.}(2023)\citenamefont
  {Kikuchi}, \citenamefont {Mc~Keever}, \citenamefont {Coopmans}, \citenamefont
  {Lubasch},\ and\ \citenamefont {Benedetti}}]{KiEtAl23}%
  \BibitemOpen
  \bibfield  {author} {\bibinfo {author} {\bibfnamefont {Y.}~\bibnamefont
  {Kikuchi}}, \bibinfo {author} {\bibfnamefont {C.}~\bibnamefont {Mc~Keever}},
  \bibinfo {author} {\bibfnamefont {L.}~\bibnamefont {Coopmans}}, \bibinfo
  {author} {\bibfnamefont {M.}~\bibnamefont {Lubasch}},\ and\ \bibinfo {author}
  {\bibfnamefont {M.}~\bibnamefont {Benedetti}},\ }\bibfield  {title} {\bibinfo
  {title} {Realization of quantum signal processing on a noisy quantum
  computer},\ }\href {https://doi.org/10.1038/s41534-023-00762-0} {\bibfield
  {journal} {\bibinfo  {journal} {Npj Quantum Inf.}\ }\textbf {\bibinfo
  {volume} {9}},\ \bibinfo {pages} {93} (\bibinfo {year} {2023})}\BibitemShut
  {NoStop}%
\bibitem [{\citenamefont {Lubasch}\ \emph {et~al.}(2018)\citenamefont
  {Lubasch}, \citenamefont {Moinier},\ and\ \citenamefont
  {Jaksch}}]{Lubasch2018}%
  \BibitemOpen
  \bibfield  {author} {\bibinfo {author} {\bibfnamefont {M.}~\bibnamefont
  {Lubasch}}, \bibinfo {author} {\bibfnamefont {P.}~\bibnamefont {Moinier}},\
  and\ \bibinfo {author} {\bibfnamefont {D.}~\bibnamefont {Jaksch}},\
  }\bibfield  {title} {\bibinfo {title} {Multigrid renormalization},\ }\href
  {https://doi.org/https://doi.org/10.1016/j.jcp.2018.06.065} {\bibfield
  {journal} {\bibinfo  {journal} {J. Comput. Phys.}\ }\textbf {\bibinfo
  {volume} {372}},\ \bibinfo {pages} {587} (\bibinfo {year}
  {2018})}\BibitemShut {NoStop}%
\bibitem [{\citenamefont {McArdle}\ \emph {et~al.}(2022)\citenamefont
  {McArdle}, \citenamefont {Gilyén},\ and\ \citenamefont
  {Berta}}]{mcardle2022quantum}%
  \BibitemOpen
  \bibfield  {author} {\bibinfo {author} {\bibfnamefont {S.}~\bibnamefont
  {McArdle}}, \bibinfo {author} {\bibfnamefont {A.}~\bibnamefont {Gilyén}},\
  and\ \bibinfo {author} {\bibfnamefont {M.}~\bibnamefont {Berta}},\
  }\href@noop {} {\bibinfo {title} {Quantum state preparation without coherent
  arithmetic}} (\bibinfo {year} {2022}),\ \Eprint
  {https://arxiv.org/abs/2210.14892} {arXiv:2210.14892 [quant-ph]} \BibitemShut
  {NoStop}%
\bibitem [{\citenamefont {Hales}\ and\ \citenamefont
  {Hallgren}(2000)}]{hales2000improved}%
  \BibitemOpen
  \bibfield  {author} {\bibinfo {author} {\bibfnamefont {L.}~\bibnamefont
  {Hales}}\ and\ \bibinfo {author} {\bibfnamefont {S.}~\bibnamefont
  {Hallgren}},\ }\bibfield  {title} {\bibinfo {title} {An improved quantum
  {Fourier} transform algorithm and applications},\ }in\ \href
  {https://doi.org/10.1109/SFCS.2000.892139} {\emph {\bibinfo {booktitle}
  {Proceedings 41st Annual Symposium on Foundations of Computer Science}}}\
  (\bibinfo {organization} {IEEE},\ \bibinfo {year} {2000})\ pp.\ \bibinfo
  {pages} {515--525}\BibitemShut {NoStop}%
\bibitem [{\citenamefont {Nam}\ \emph {et~al.}(2020)\citenamefont {Nam},
  \citenamefont {Su},\ and\ \citenamefont {Maslov}}]{nam2020approximate}%
  \BibitemOpen
  \bibfield  {author} {\bibinfo {author} {\bibfnamefont {Y.}~\bibnamefont
  {Nam}}, \bibinfo {author} {\bibfnamefont {Y.}~\bibnamefont {Su}},\ and\
  \bibinfo {author} {\bibfnamefont {D.}~\bibnamefont {Maslov}},\ }\bibfield
  {title} {\bibinfo {title} {Approximate quantum {Fourier} transform with {O}(n
  log(n)) {T} gates},\ }\href {https://doi.org/10.1038/s41534-020-0257-5}
  {\bibfield  {journal} {\bibinfo  {journal} {npj Quantum Inf.}\ }\textbf
  {\bibinfo {volume} {6}},\ \bibinfo {pages} {26} (\bibinfo {year}
  {2020})}\BibitemShut {NoStop}%
\bibitem [{\citenamefont {Griffiths}\ and\ \citenamefont
  {Niu}(1996)}]{griffiths1996semiclassical}%
  \BibitemOpen
  \bibfield  {author} {\bibinfo {author} {\bibfnamefont {R.~B.}\ \bibnamefont
  {Griffiths}}\ and\ \bibinfo {author} {\bibfnamefont {C.-S.}\ \bibnamefont
  {Niu}},\ }\bibfield  {title} {\bibinfo {title} {{Semiclassical Fourier
  Transform for Quantum Computation}},\ }\href
  {https://doi.org/10.1103/PhysRevLett.76.3228} {\bibfield  {journal} {\bibinfo
   {journal} {Phys. Rev. Lett.}\ }\textbf {\bibinfo {volume} {76}},\ \bibinfo
  {pages} {3228} (\bibinfo {year} {1996})}\BibitemShut {NoStop}%
\bibitem [{\citenamefont {Chiaverini}\ \emph {et~al.}(2005)\citenamefont
  {Chiaverini}, \citenamefont {Britton}, \citenamefont {Leibfried},
  \citenamefont {Knill}, \citenamefont {Barrett}, \citenamefont {Blakestad},
  \citenamefont {Itano}, \citenamefont {Jost}, \citenamefont {Langer},
  \citenamefont {Ozeri}, \citenamefont {Schaetz},\ and\ \citenamefont
  {Wineland}}]{chiaverini2005implementation}%
  \BibitemOpen
  \bibfield  {author} {\bibinfo {author} {\bibfnamefont {J.}~\bibnamefont
  {Chiaverini}}, \bibinfo {author} {\bibfnamefont {J.}~\bibnamefont {Britton}},
  \bibinfo {author} {\bibfnamefont {D.}~\bibnamefont {Leibfried}}, \bibinfo
  {author} {\bibfnamefont {E.}~\bibnamefont {Knill}}, \bibinfo {author}
  {\bibfnamefont {M.~D.}\ \bibnamefont {Barrett}}, \bibinfo {author}
  {\bibfnamefont {R.~B.}\ \bibnamefont {Blakestad}}, \bibinfo {author}
  {\bibfnamefont {W.~M.}\ \bibnamefont {Itano}}, \bibinfo {author}
  {\bibfnamefont {J.~D.}\ \bibnamefont {Jost}}, \bibinfo {author}
  {\bibfnamefont {C.}~\bibnamefont {Langer}}, \bibinfo {author} {\bibfnamefont
  {R.}~\bibnamefont {Ozeri}}, \bibinfo {author} {\bibfnamefont
  {T.}~\bibnamefont {Schaetz}},\ and\ \bibinfo {author} {\bibfnamefont {D.~J.}\
  \bibnamefont {Wineland}},\ }\bibfield  {title} {\bibinfo {title}
  {{Implementation of the Semiclassical Quantum Fourier Transform in a Scalable
  System}},\ }\href {https://doi.org/10.1126/science.1110335} {\bibfield
  {journal} {\bibinfo  {journal} {Science}\ }\textbf {\bibinfo {volume}
  {308}},\ \bibinfo {pages} {997} (\bibinfo {year} {2005})}\BibitemShut
  {NoStop}%
\bibitem [{\citenamefont {Lubinski}\ \emph {et~al.}(2023)\citenamefont
  {Lubinski}, \citenamefont {Johri}, \citenamefont {Varosy}, \citenamefont
  {Coleman}, \citenamefont {Zhao}, \citenamefont {Necaise}, \citenamefont
  {Baldwin}, \citenamefont {Mayer},\ and\ \citenamefont
  {Proctor}}]{lubinski2023applicationoriented}%
  \BibitemOpen
  \bibfield  {author} {\bibinfo {author} {\bibfnamefont {T.}~\bibnamefont
  {Lubinski}}, \bibinfo {author} {\bibfnamefont {S.}~\bibnamefont {Johri}},
  \bibinfo {author} {\bibfnamefont {P.}~\bibnamefont {Varosy}}, \bibinfo
  {author} {\bibfnamefont {J.}~\bibnamefont {Coleman}}, \bibinfo {author}
  {\bibfnamefont {L.}~\bibnamefont {Zhao}}, \bibinfo {author} {\bibfnamefont
  {J.}~\bibnamefont {Necaise}}, \bibinfo {author} {\bibfnamefont {C.~H.}\
  \bibnamefont {Baldwin}}, \bibinfo {author} {\bibfnamefont {K.}~\bibnamefont
  {Mayer}},\ and\ \bibinfo {author} {\bibfnamefont {T.}~\bibnamefont
  {Proctor}},\ }\href@noop {} {\bibinfo {title} {Application-oriented
  performance benchmarks for quantum computing}} (\bibinfo {year} {2023}),\
  \Eprint {https://arxiv.org/abs/2110.03137} {arXiv:2110.03137 [quant-ph]}
  \BibitemShut {NoStop}%
\bibitem [{\citenamefont {Abdulsatar}(2021)}]{abdulsatar2021asic}%
  \BibitemOpen
  \bibfield  {author} {\bibinfo {author} {\bibfnamefont {A.~A.}\ \bibnamefont
  {Abdulsatar}},\ }\bibfield  {title} {\bibinfo {title} {Asic implementation of
  high-speed vector magnitude \& arctangent approximator},\ }\href
  {https://doi.org/10.18721/JCSTCS.14401} {\bibfield  {journal} {\bibinfo
  {journal} {Computing, Telecommunications and Control}\ }\textbf {\bibinfo
  {volume} {14}},\ \bibinfo {pages} {7} (\bibinfo {year} {2021})}\BibitemShut
  {NoStop}%
\bibitem [{\citenamefont {Cerezo}\ \emph {et~al.}(2021)\citenamefont {Cerezo},
  \citenamefont {Arrasmith}, \citenamefont {Babbush}, \citenamefont {Benjamin},
  \citenamefont {Endo}, \citenamefont {Fujii}, \citenamefont {McClean},
  \citenamefont {Mitarai}, \citenamefont {Yuan}, \citenamefont {Cincio},\ and\
  \citenamefont {Coles}}]{Cerezo2021}%
  \BibitemOpen
  \bibfield  {author} {\bibinfo {author} {\bibfnamefont {M.}~\bibnamefont
  {Cerezo}}, \bibinfo {author} {\bibfnamefont {A.}~\bibnamefont {Arrasmith}},
  \bibinfo {author} {\bibfnamefont {R.}~\bibnamefont {Babbush}}, \bibinfo
  {author} {\bibfnamefont {S.~C.}\ \bibnamefont {Benjamin}}, \bibinfo {author}
  {\bibfnamefont {S.}~\bibnamefont {Endo}}, \bibinfo {author} {\bibfnamefont
  {K.}~\bibnamefont {Fujii}}, \bibinfo {author} {\bibfnamefont {J.~R.}\
  \bibnamefont {McClean}}, \bibinfo {author} {\bibfnamefont {K.}~\bibnamefont
  {Mitarai}}, \bibinfo {author} {\bibfnamefont {X.}~\bibnamefont {Yuan}},
  \bibinfo {author} {\bibfnamefont {L.}~\bibnamefont {Cincio}},\ and\ \bibinfo
  {author} {\bibfnamefont {P.~J.}\ \bibnamefont {Coles}},\ }\bibfield  {title}
  {\bibinfo {title} {Variational quantum algorithms},\ }\href
  {https://doi.org/10.1038/s42254-021-00348-9} {\bibfield  {journal} {\bibinfo
  {journal} {Nat. Rev. Phys.}\ }\textbf {\bibinfo {volume} {3}},\ \bibinfo
  {pages} {625} (\bibinfo {year} {2021})}\BibitemShut {NoStop}%
\bibitem [{\citenamefont {Liu}\ and\ \citenamefont
  {Nocedal}(1989)}]{liu1989limited}%
  \BibitemOpen
  \bibfield  {author} {\bibinfo {author} {\bibfnamefont {D.~C.}\ \bibnamefont
  {Liu}}\ and\ \bibinfo {author} {\bibfnamefont {J.}~\bibnamefont {Nocedal}},\
  }\bibfield  {title} {\bibinfo {title} {On the limited memory {BFGS} method
  for large scale optimization},\ }\href {https://doi.org/10.1007/BF01589116}
  {\bibfield  {journal} {\bibinfo  {journal} {Math. Program.}\ }\textbf
  {\bibinfo {volume} {45}},\ \bibinfo {pages} {503} (\bibinfo {year}
  {1989})}\BibitemShut {NoStop}%
\bibitem [{\citenamefont {McClean}\ \emph {et~al.}(2018)\citenamefont
  {McClean}, \citenamefont {Boixo}, \citenamefont {Smelyanskiy}, \citenamefont
  {Babbush},\ and\ \citenamefont {Neven}}]{McClean2018}%
  \BibitemOpen
  \bibfield  {author} {\bibinfo {author} {\bibfnamefont {J.~R.}\ \bibnamefont
  {McClean}}, \bibinfo {author} {\bibfnamefont {S.}~\bibnamefont {Boixo}},
  \bibinfo {author} {\bibfnamefont {V.~N.}\ \bibnamefont {Smelyanskiy}},
  \bibinfo {author} {\bibfnamefont {R.}~\bibnamefont {Babbush}},\ and\ \bibinfo
  {author} {\bibfnamefont {H.}~\bibnamefont {Neven}},\ }\bibfield  {title}
  {\bibinfo {title} {Barren plateaus in quantum neural network training
  landscapes},\ }\href {https://doi.org/10.1038/s41467-018-07090-4} {\bibfield
  {journal} {\bibinfo  {journal} {Nat. Commun.}\ }\textbf {\bibinfo {volume}
  {9}},\ \bibinfo {pages} {4812} (\bibinfo {year} {2018})}\BibitemShut
  {NoStop}%
\bibitem [{\citenamefont {Plekhanov}\ \emph {et~al.}(2022)\citenamefont
  {Plekhanov}, \citenamefont {Rosenkranz}, \citenamefont {Fiorentini},\ and\
  \citenamefont {Lubasch}}]{Plekhanov2022variationalquantum}%
  \BibitemOpen
  \bibfield  {author} {\bibinfo {author} {\bibfnamefont {K.}~\bibnamefont
  {Plekhanov}}, \bibinfo {author} {\bibfnamefont {M.}~\bibnamefont
  {Rosenkranz}}, \bibinfo {author} {\bibfnamefont {M.}~\bibnamefont
  {Fiorentini}},\ and\ \bibinfo {author} {\bibfnamefont {M.}~\bibnamefont
  {Lubasch}},\ }\bibfield  {title} {\bibinfo {title} {Variational quantum
  amplitude estimation},\ }\href {https://doi.org/10.22331/q-2022-03-17-670}
  {\bibfield  {journal} {\bibinfo  {journal} {{Quantum}}\ }\textbf {\bibinfo
  {volume} {6}},\ \bibinfo {pages} {670} (\bibinfo {year} {2022})}\BibitemShut
  {NoStop}%
\bibitem [{\citenamefont {Marin-Sanchez}\ \emph {et~al.}(2023)\citenamefont
  {Marin-Sanchez}, \citenamefont {Gonzalez-Conde},\ and\ \citenamefont
  {Sanz}}]{marin2023quantum}%
  \BibitemOpen
  \bibfield  {author} {\bibinfo {author} {\bibfnamefont {G.}~\bibnamefont
  {Marin-Sanchez}}, \bibinfo {author} {\bibfnamefont {J.}~\bibnamefont
  {Gonzalez-Conde}},\ and\ \bibinfo {author} {\bibfnamefont {M.}~\bibnamefont
  {Sanz}},\ }\bibfield  {title} {\bibinfo {title} {Quantum algorithms for
  approximate function loading},\ }\href
  {https://doi.org/10.1103/PhysRevResearch.5.033114} {\bibfield  {journal}
  {\bibinfo  {journal} {Phys. Rev. Res.}\ }\textbf {\bibinfo {volume} {5}},\
  \bibinfo {pages} {033114} (\bibinfo {year} {2023})}\BibitemShut {NoStop}%
\bibitem [{\citenamefont {Akhalwaya}\ \emph {et~al.}(2023)\citenamefont
  {Akhalwaya}, \citenamefont {Connolly}, \citenamefont {Guichard},
  \citenamefont {Herbert}, \citenamefont {Kargi}, \citenamefont {Krajenbrink},
  \citenamefont {Lubasch}, \citenamefont {{Mc Keever}}, \citenamefont {Sorci},
  \citenamefont {Spranger},\ and\ \citenamefont
  {Williams}}]{akhalwaya2023modular}%
  \BibitemOpen
  \bibfield  {author} {\bibinfo {author} {\bibfnamefont {I.~Y.}\ \bibnamefont
  {Akhalwaya}}, \bibinfo {author} {\bibfnamefont {A.}~\bibnamefont {Connolly}},
  \bibinfo {author} {\bibfnamefont {R.}~\bibnamefont {Guichard}}, \bibinfo
  {author} {\bibfnamefont {S.}~\bibnamefont {Herbert}}, \bibinfo {author}
  {\bibfnamefont {C.}~\bibnamefont {Kargi}}, \bibinfo {author} {\bibfnamefont
  {A.}~\bibnamefont {Krajenbrink}}, \bibinfo {author} {\bibfnamefont
  {M.}~\bibnamefont {Lubasch}}, \bibinfo {author} {\bibfnamefont
  {C.}~\bibnamefont {{Mc Keever}}}, \bibinfo {author} {\bibfnamefont
  {J.}~\bibnamefont {Sorci}}, \bibinfo {author} {\bibfnamefont
  {M.}~\bibnamefont {Spranger}},\ and\ \bibinfo {author} {\bibfnamefont
  {I.}~\bibnamefont {Williams}},\ }\href@noop {} {\bibinfo {title} {{A Modular
  Engine for Quantum Monte Carlo Integration}}} (\bibinfo {year} {2023}),\
  \Eprint {https://arxiv.org/abs/2308.06081} {arXiv:2308.06081 [quant-ph]}
  \BibitemShut {NoStop}%
\bibitem [{\citenamefont {Sivarajah}\ \emph {et~al.}(2020)\citenamefont
  {Sivarajah}, \citenamefont {Dilkes}, \citenamefont {Cowtan}, \citenamefont
  {Simmons}, \citenamefont {Edgington},\ and\ \citenamefont
  {Duncan}}]{sivarajah2020t}%
  \BibitemOpen
  \bibfield  {author} {\bibinfo {author} {\bibfnamefont {S.}~\bibnamefont
  {Sivarajah}}, \bibinfo {author} {\bibfnamefont {S.}~\bibnamefont {Dilkes}},
  \bibinfo {author} {\bibfnamefont {A.}~\bibnamefont {Cowtan}}, \bibinfo
  {author} {\bibfnamefont {W.}~\bibnamefont {Simmons}}, \bibinfo {author}
  {\bibfnamefont {A.}~\bibnamefont {Edgington}},\ and\ \bibinfo {author}
  {\bibfnamefont {R.}~\bibnamefont {Duncan}},\ }\bibfield  {title} {\bibinfo
  {title} {t|ket⟩: a retargetable compiler for nisq devices},\ }\href
  {https://doi.org/10.1088/2058-9565/ab8e92} {\bibfield  {journal} {\bibinfo
  {journal} {Quantum Sci. Technol.}\ }\textbf {\bibinfo {volume} {6}},\
  \bibinfo {pages} {014003} (\bibinfo {year} {2020})}\BibitemShut {NoStop}%
\bibitem [{\citenamefont {An}\ \emph {et~al.}(2023)\citenamefont {An},
  \citenamefont {Liu}, \citenamefont {Wang},\ and\ \citenamefont
  {Zhao}}]{an2023theory}%
  \BibitemOpen
  \bibfield  {author} {\bibinfo {author} {\bibfnamefont {D.}~\bibnamefont
  {An}}, \bibinfo {author} {\bibfnamefont {J.-P.}\ \bibnamefont {Liu}},
  \bibinfo {author} {\bibfnamefont {D.}~\bibnamefont {Wang}},\ and\ \bibinfo
  {author} {\bibfnamefont {Q.}~\bibnamefont {Zhao}},\ }\href@noop {} {\bibinfo
  {title} {A theory of quantum differential equation solvers: limitations and
  fast-forwarding}} (\bibinfo {year} {2023}),\ \Eprint
  {https://arxiv.org/abs/2211.05246} {arXiv:2211.05246 [quant-ph]} \BibitemShut
  {NoStop}%
\end{thebibliography}%

\end{document}